\documentclass[journal]{IEEEtran}
\usepackage{graphicx}
\usepackage{epsfig}
\usepackage{latexsym}
\usepackage{amsfonts}
\usepackage{here}
\usepackage{rawfonts}
\usepackage[T1]{fontenc}
\usepackage{calc}
\usepackage{url}
\usepackage{enumerate}
\usepackage{color}
\usepackage[tbtags]{amsmath}
\usepackage{amssymb}
\usepackage{upref}
\usepackage{epic,eepic}
\usepackage{times}
\usepackage{dsfont}
\usepackage{comment}
\usepackage{cite}

\usepackage{graphicx}
\usepackage[font={small}]{caption}
\usepackage[font={small}]{subcaption}

\usepackage{stfloats}
\usepackage{mathtools}
\usepackage{amsmath}

\newtheorem{theorem}{\bf{Theorem}}
\newtheorem{definition}{\bf{Definition}}

\ifCLASSINFOpdf
\else
\fi

\begin{document}

\title{Stochastic Geometry Modeling and Analysis of Finite Millimeter Wave Wireless Networks}
\author{Seyed~Mohammad~Azimi-Abarghouyi,~Behrooz~Makki,~\\Masoumeh Nasiri-Kenari,~\IEEEmembership{Senior~Member,~IEEE}, and~Tommy~Svensson,~\IEEEmembership{Senior~Member,~IEEE}
\thanks{S.M. Azimi-Abarghouyi and M. Nasiri-Kenari are with the Dep. of Electrical Engineering, Sharif
University of Technology, Tehran 11365-9363, Iran. (e-mail: azimi$\_$sm@ee.sharif.edu; mnasiri@sharif.edu). B. Makki and T. Svensson are with the Dep. of Electrical Engineering, Chalmers University of Technology, 412 96 Gothenburg, Sweden. (e-mail: $\{$behrooz.makki, tommy.svensson$\}$@chalmers.se). This work has been supported by the Research Office of Sharif University of Technology under grant QB960605, Iran, and by the VR Research Link grant, Sweden.}
}

\maketitle

\begin{abstract}
This paper develops a stochastic geometry-based approach for the modeling and analysis of finite millimeter wave (mmWave) wireless networks where a random number of transmitters and receivers are randomly located inside a finite region. We consider a selection strategy to serve a reference receiver by the transmitter providing the maximum average received power among all transmitters. Considering the unique features of mmWave communications such as directional transmit and receive beamforming and having different channels for line-of-sight (LOS) and non-line-of-sight (NLOS) links according to the blockage process, we study the coverage probability and the ergodic rate for the reference receiver that can be located everywhere inside the network region. As key steps for the analyses, the distribution of the distance from the reference receiver to its serving LOS or NLOS transmitter and LOS and NLOS association probabilities are derived. We also derive the Laplace transform of the interferences from LOS and NLOS transmitters. Finally, we propose upper and lower bounds on the coverage probability that can be evaluated easier than the exact results, and investigate the impact of different parameters including the receiver location, the beamwidth, and the blockage process exponent on the system performance. 
\end{abstract}
\section{Introduction}
Millimeter wave (mmWave) communications is a promising candidate technology for the next generation of wireless networks [1]. This is mainly because mmWave frequencies provide large bandwidth, compatibility with directional communications, and possibility of dense deployments. However, the signal propagation at mmWave frequencies suffers from poor penetration, diffraction and scattering through blockages [2]-[3]. On the other hand, the ever-growing randomness and irregularity in the locations of nodes in a wireless network has led to a growing interest in the use of stochastic geometry and Poisson point processes (PPPs) for accurate and tractable spatial modeling and analysis [4]-[6]. In this way, based on the proposed models for the directionality of antennas and blockage process in [7]-[8], most works exploit infinite homogeneous PPP (HPPP) [6, Def. 2.8] to model and analyze the performance of different mmWave wireless networks over an infinite region [9]-[14]. However, in practice wireless networks do not spread over an infinite region. Moreover, deployments of mmWave wireless networks over small finite regions are becoming mainstream, thanks to the popularity of mmWave in short-range communications, indoor and ad hoc networks such as WirelessHD and IEEE 802.11ad standards [15]-[17]. 

The modeling and performance analysis of finite wireless networks are more challenging and require different approaches in comparison to infinite wireless networks, even in microwave frequencies with no beamforming and blockage effects [18]-[20]. The main challenge is that a finite point process is not statistically similar from different locations, and therefore, the system performance depends on the receiver location [18]. Finite mmWave wireless networks have been mostly studied based on the binomial point process (BPP) [6, Def. 2.11], where a fixed and finite number of nodes are distributed independently and uniformly inside a finite region. Considering the BPP, the state-of-the-art works are focused on wearable device-to-device applications and present performance characterizations of a fixed link inside a finite region of people who are considered both as interferers and blockages [21]-[23]. Although fixed-link analysis provides useful insights for the performance of device-to-device use-case scenarios, it is not suitable for networks with infrastructure such as cellular networks that can serve a receiver by a transmitter with the highest quality performance.

In this paper, we provide a tractable model for finite mmWave wireless networks using the finite homogeneous Poisson point process (FHPPP) proposed in [18, Def. 1], which is a suitable point process to model a random number of nodes randomly located inside a finite region. We consider a transmitter selection strategy referred to as average received power selection, where a reference receiver is served by the transmitter with the maximum received power averaged over small-scale fading in the network. We derive the coverage probability and the ergodic rate of the reference receiver under the considered selection strategy and mmWave  features including directional transmit and receive beamforming and different line-of-sight (LOS) and non-line-of-sight (NLOS) link characteristics. As key steps for the coverage probability and the ergodic rate analyses, the distribution of the distance from the reference receiver to its serving transmitter, association probabilities, and the Laplace transform (LT) of the interference are derived for both sets of LOS and NLOS transmitters from the reference receiver. As a part of the LT of the interference derivation, the distribution of the overall transmit and receive gain of a link is characterized. We also propose lower and upper bounds on the coverage probability that are more computationally tractable results.

We investigate the impact of different parameters of the system model on the coverage probability and the ergodic rate. Our analysis reveals that there exists a blockage exponent that maximizes the coverage probability. Also, there is an optimal distance for the location of the reference receiver from the center of the network in terms of the coverage probability and the ergodic rate. As another observed trend, increasing the transmit and receive antenna beamwidths decreases the coverage probability. Our evaluations also show that our proposed upper bound for a small antenna beamwidth and our proposed lower bound for a large antenna beamwidth tightly mimic the exact results on the coverage probability.

Our work is different from the state-of-the-art literature, e.g., [21]-[23], from two perspectives. First, different from the BPP which models a fixed number of nodes in a region, we consider the FHPPP [18], which is suitable for finite regions with a random number of nodes, and comprehensively address the modeling and analysis of finite mmWave wireless networks using the properties of the PPP. In this regard, we perform new analyses considering a new system model and new assumptions. Second, we consider a transmitter selection strategy that provides the maximum averaged received power in the allocation of a transmitter to a receiver, as assumed also in previous works on infinite mmWave networks [8]-[12]. 

The rest of the paper is organized as follows. Section II describes the system model and the selection strategy. Section III characterizes the link distance distributions and the association probabilities. Section IV presents the analytical results for the coverage probability and the ergodic rate of finite mmwave wireless networks and derives the LT of the interference as well as upper and lower bounds on the coverage probability. Section V presents the numerical and simulation results. Finally, Section VI concludes the paper.

\section{System Model}
In this section, we provide a mathematical model of the system. We begin with the spatial distribution of the nodes. Then, we describe the channel model and the transmitter selection strategy.

\subsection{Spatial Model}
We consider a finite mmWave wireless network as shown in Fig. 1. The locations of transmitters are modeled as an FHPPP ${\Phi}_\text{T}$ with intensity $\lambda_\text{T}$ over a finite region ${\mathbf{\cal A}} \subset \mathbb{R}^2$, which is defined in the following.
\begin{definition}
The FHPPP is defined as ${\Phi}=\mathcal{P}\cap{\cal A}$, where $\mathcal{P}$ is an HPPP of intensity $\lambda$ and ${\cal A}\subset\mathbb{R}^2$ [18].  \hspace{395pt}\IEEEQEDclosed
\end{definition}

Receivers are also located inside $\cal A$ according to another FHPPP ${\Phi}_\text{R}$ with intensity $\lambda_\text{R}$ that is independent of ${\Phi}_\text{T}$. We assume that $\lambda_\text{R}\gg\lambda_\text{T}$ and the transmitters are all active and transmit at the same power. In each of the available resource blocks, each transmitter is assumed to serve a single receiver that is randomly selected among its associated receivers. Then, the intensity of active receivers in each resource block, denoted by a point process ${\hat \Phi}_\text{R}$, is equal to $\lambda_\text{T}$. 
\begin{figure}[tb!]
\centering

\includegraphics[width =2.5in]{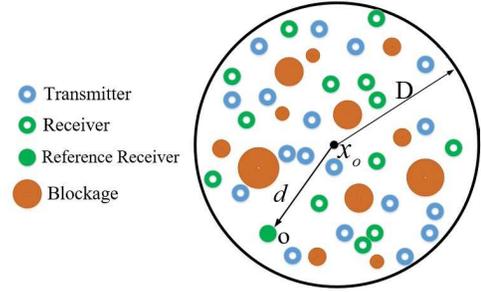}

\caption{An illustration of the system model for finite mmWave wireless networks.}
\vspace{-10pt}
\end{figure}

As the signal propagation at mmWave frequencies suffers from poor penetration, a link is LOS or NLOS depending on whether or not it is intersected by a blockage. In harmony with, e.g., [8]-[14], [22], we assume that there is no correlation in the blockage process such that a link with length $r$ is LOS with probability $p_\text{L}(r)$ or NLOS with probability $p_\text{N}(r)=1-p_\text{L}(r)$. As a result from the location of a receiver, the transmitters can be split into two independent tiers comprising a finite non-homogeneous Poisson point process (FNPPP) ${\Phi}_\text{L}$ with intensity $\lambda_\text{T}p_\text{L}(r)$ for LOS transmitters and an FNPPP ${\Phi}_\text{N}$ with intensity $\lambda_\text{T}(1-p_\text{L}(r))$ for NLOS transmitters, such that ${\Phi}_\text{T}={\Phi}_\text{L}\cup{\Phi}_\text{N}$. We also denote the number of LOS and NLOS transmitters by $n_\text{L}$ and $n_\text{N}$, respectively. The FNPPP is defined as follows. 
\begin{definition}
We define an FNPPP ${\Phi}$ with the non-constant intensity function $\lambda(\mathbf{z})$ at a location $\mathbf{z}$ over ${\mathbf{\cal A}} \subset \mathbb{R}^2$ such that the probability that $n$ points are in a region ${\mathbf{\cal B}} \subset \mathbb{R}^2$ is given by
\begin{eqnarray}
{\mathbb{P}}\left( {{\Phi} \left( \mathbf{\cal B} \right) = n} \right) = \exp\left( -{\Lambda}({\rm{\cal C}})\right)\frac{{\Lambda}({\rm{\cal C}})^n}{{n!}},
\end{eqnarray}
where ${\Lambda}({\rm{\cal C}}) = \int_{\cal C}^{} \lambda (\mathbf{z})\mathrm{d}\mathbf{z}$ is the intensity measure and ${\rm{\mathbf{\cal C}}}$ denotes the intersection between $\rm{\mathbf{\cal A}}$ and $\rm{\mathbf{\cal B}}$, i.e., $\rm{\mathbf{\cal C}} = \cal A \cap \cal B$.\hspace{370pt}\IEEEQEDclosed
\end{definition}

For simplicity and in harmony with, e.g., [18]-[20], we let ${\mathbf{\cal A}}=\mathbf{b}(\mathbf{x}_\mathbf{o},D)$, where $\mathbf{b}(\mathbf{x}_\mathbf{o},D)$ represents a disk centered at $\mathbf{x}_\mathbf{o}$ with radius $D$. However, our theoretical results can be extended to the case of an arbitrarily-shaped region ${\mathbf{\cal A}}$. 

Receivers can be located everywhere in ${\mathbf{\cal A}}$. With no loss of generality, we conduct the analysis for a reference receiver located at the origin $\mathbf{o}$. We further define $d = \|\mathbf{x}_\mathbf{o}\|$, which denotes the distance from the reference receiver to the center of $\cal A$, i.e., $\mathbf{x}_\mathbf{o}$.

\subsection{Channel Model}
As the LOS and NLOS propagation have different characteristics, we consider the received power at the reference receiver from a transmitter located at $\mathbf{y} \in {\Phi}_\text{T}$ as
${h_\mathbf{y}}{G_\mathbf{y}}{\| \mathbf{y} \|^{ - \alpha_q }},\ q = \left\{\text{L},\text{N}\right\}$,
where $\alpha_\text{L}$ and $\alpha_\text{N}$ are the pathloss exponents for the LOS and NLOS links, respectively. Note that the NLOS mmWave signals typically exhibit a higher pathloss exponent, i.e., $\alpha_\text{L}<\alpha_\text{N}$. Assuming independent Nakagami fading for each link, the fading power $h_\mathbf{y}$ can be modeled as a normalized Gamma random variable ${\Gamma}(v_\text{L}, \frac{1}{v_\text{L}})$ if the link is LOS and ${\Gamma}(v_\text{N}, \frac{1}{v_\text{N}})$ if the link is NLOS. Also, ${G_\mathbf{y}}$ denotes the overall antenna gain. Also, we consider the same closed-in reference distance for both LOS and NLOS links to have same intercepts [8], [24]. 

\begin{figure}[tb!]
\centering

\includegraphics[width =1.5in]{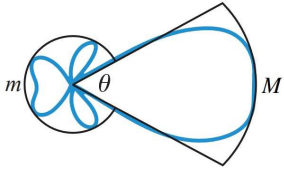}

\caption{Approximated sectored-pattern antenna model.}

\end{figure}
To compensate for high propagation losses, mmWave transmitters and receivers use large antenna arrays to communicate directionaly. We assume the approximated sectored-pattern antenna model in Fig. 2, according to which the transmitter gain $G_\text{T}$ and the receiver gain $G_\text{R}$ can be given by
\begin{eqnarray}
G_q (\theta) = \left\{ {\begin{array}{*{20}{c}}
{\hspace{-6pt}M_q \hspace{+30pt} |\theta|<\frac{1}{2}\theta_q,}\\
{\hspace{-6pt}m_q \hspace{+33pt} \text{otherwise},\hspace{-5pt}}
\end{array}} \right. , \ q = \left\{\text{T},\text{R}\right\},
\end{eqnarray}
where $\theta$ denotes the angle in polar coordinates, $\theta_q$ is the beamwidth, and $M_q$ and $m_q$ are the main-lobe and side-lobe gain, respectively, i.e., $M_q>m_q$. Therefore, the overall antenna gain $G_\mathbf{y}$ which is equal to $G_\text{T} \times G_\text{R}$ can be one of $\left\{M_\text{T}M_\text{R},M_\text{T}m_\text{R},m_\text{T}M_\text{R},m_\text{T}m_\text{R}\right\}$ according to directions of the transmitter and the receiver of the link. For notational simplicity, let us use the auxiliary variables $a_1 = M_\text{T}M_\text{R}$, $a_2 = M_\text{T}m_\text{R}$, $a_3 = m_\text{T}M_\text{R}$, and $a_4 = m_\text{T}m_\text{R}$.

\subsection{Selection Strategy}
We assume average received power selection strategy where a reference receiver is associated to the transmitter that provides the maximum received power averaged over the fading. Therefore, the candidate among the LOS transmitters is the closest one and found as $\mathbf{x}_\text{L} = \arg \mathop {\min }\limits_{{\mathbf{x}} \in \left\{{\Phi}_\text{L}\mid n_\text{L}\geq 1\right\}}\|\mathbf{x}\|$, while among the NLOS transmitters the candidate is found as $\mathbf{x}_\text{N} = \arg \mathop {\min }\limits_{{\mathbf{x}} \in \left\{{\Phi}_\text{N}\mid n_\text{N}\geq 1\right\}}\|\mathbf{x}\|$. Finally, the serving transmitter is selected between the LOS and NLOS candidates as
\begin{align}
\mathbf{x}_q = \left\{\begin{matrix}
\mathbf{x}_\text{L} \hspace{0pt} &\text{if}  \ n_\text{L}\geq 1\hspace{+2pt}\&\hspace{+2pt}n_\text{N}=0,\\ 
\mathbf{x}_\text{N} \hspace{0pt}&\text{if}  \ n_\text{L}=0\hspace{+2pt}\&\hspace{+2pt}n_\text{N}\geq 1,\\ 
\arg \max \left\{\|\mathbf{x}_\text{L}\|^{-{\alpha}_\text{L}},\|\mathbf{x}_\text{N}\|^{-{\alpha}_\text{N}}\right\}\hspace{0pt}& \text{if}  \ n_\text{L}\geq 1\hspace{+2pt}\&\hspace{+2pt}n_\text{N}\geq 1.
 \end{matrix}\right.
\end{align}
Assuming that the main antenna beams of the serving transmitter and the reference receiver are aligned for the maximum overall antenna gain, i.e., $a_1$,\footnote{Such alignment can be performed by sophisticated beam training protocols [2].} the signal-to-interference-and-noise ratio (SINR) at the origin can be expressed as
\begin{eqnarray}
\text{SINR}_q = \frac{{{a_1}{h_q}{{\| {{\mathbf{x}_q}} \|}^{ - {\alpha _q}}}}}{{{\sigma ^2} + {{\cal I}_\text{L}} + {{\cal I}_\text{N}}}}, \ q = \left\{\text{L},\text{N}\right\},
\end{eqnarray}
where $\sigma ^2$ is the noise power, and ${\cal I}_\text{L} = \mathop \sum \limits_{\mathbf{y} \in {\Phi}_\text{L} \backslash \left\{{\mathbf{x}_q}\right\} } {G_\mathbf{y}}{h_\mathbf{y}}{\| \mathbf{y} \|^{ - {\alpha _\text{L}}}}$ and ${\cal I}_\text{N} = \mathop \sum \limits_{\mathbf{y} \in {\Phi}_\text{N} \backslash  \left\{{\mathbf{x}_q}\right\} } {G_\mathbf{y}}{h_\mathbf{y}}{\| \mathbf{y} \|^{ - {\alpha _\text{N}}}}$ are the interferences from LOS and NLOS transmitters, respectively. 

\section{Association Probability and Serving Distance Distribution}
This section derives the probability that a reference receiver with distance $d$ to the center of $\cal A$ is served by a given LOS or NLOS tier of transmitters, which is termed as the association
probability. Then, we derive the distribution of the distance from the reference receiver to its serving transmitter depending on the association to an LOS or NLOS transmitter. These association probabilities and distance distributions are used later in the coverage probability and the ergodic rate analyses. 

According to (3), in order to present the results, we first need to derive the distance distributions of the reference receiver to its closest LOS and NLOS transmitters.

The distance from the reference receiver to its closest LOS transmitter, i.e., ${\| \mathbf{x}_\text{L} \|}$, is larger than $r$ if and only if at least one transmitter exists inside $\cal A$ and there is no transmitter located within intersection $\mathbf{b}(\mathbf{o},r)\cap \cal A$. Letting ${\rm{\cal C}_{r}}$ denote the intersection, we have
\begin{align}
{\mathbb{P}}\left( {{\| \mathbf{x}_\text{L} \|} > r}\right) &=\frac{\mathbb{P}(n(\Phi_\text{L}\cap {\cal C}_r)=0 \hspace{+2pt}\& \hspace{+2pt}n_\text{L}\geq 1)}{\mathbb{P}(n_\text{L}\geq 1)} \nonumber\hspace{+0pt}\\&\stackrel{(a)}{=}\frac{\mathbb{P}(n(\Phi_\text{L}\cap {\cal C}_r)=0) \mathbb{P}(n(\Phi_\text{L}\backslash {\cal C}_r)\geq 1)}{\mathbb{P}(n_\text{L}\geq 1)} \nonumber\hspace{0pt}\\&\stackrel{(b)}{=} \frac{{\rm{exp}}\left( -{\Lambda}_\text{L}({\rm{\cal C}_r}) \right)(1-{\rm{exp}}\left( -{\Lambda}_\text{L}({{\cal A}\backslash \rm{{\cal C}_r}}) \right))}{1-{\rm{exp}}\left( -{\Lambda}_\text{L}({\rm{\cal A}}) \right)} \nonumber\\&= \frac{{\rm{exp}}\left( -{\Lambda}_\text{L}({\rm{\cal C}_r}) \right)-{\rm{exp}}\left( -{\Lambda}_\text{L}({\rm{\cal A}}) \right)}{1-{\rm{exp}}\left( -{\Lambda}_\text{L}({\rm{\cal A}}) \right)},\hspace{0pt}
\end{align}
where $(a)$ follows from the fact that the numbers of points of a PPP in disjoint regions are independent, and $(b)$ is due to the fact that $\|\mathbf{x}_\text{L}\|\leq D+d$ and ${\Lambda}_\text{L}({\rm{\cal C}_r})+{\Lambda}_\text{L}({{\cal A}\backslash \rm{{\cal C}_r}}) = {\Lambda}_\text{L}({{\cal A}}) $. Note that when the intersection is the whole of $\cal A$, (5) becomes zero. 

To convert from Cartesian to polar coordinates, (5) can be obtained according to the following cases.




\textbf{Case 1:} If ${\cal A}\cap \mathbf{b}(\mathbf{o},r) = \mathbf{b}(\mathbf{o},r)$, i.e., $0\leq r<D-d$, then
\begin{eqnarray}
{\Lambda}_\text{L}({\rm{\cal C}_r}) = \int_{ 0}^{2\pi}\int_{0}^{r}\lambda_\text{T} p_\text{L}(x)x\mathrm{d}x\mathrm{d}\theta = 2\pi \lambda_\text{T} \int_{0}^{r} xp_\text{L}(x) \mathrm{d}x.
\end{eqnarray}

\textbf{Case 2:} If ${\cal A}\cap \mathbf{b}(\mathbf{o},r) \ne \mathbf{b}(\mathbf{o},r)$, i.e., $D-d \leq r<D+d$, then
\begin{align}
&{\Lambda}_\text{L}({\rm{\cal C}_r}) =  \int_{ -\varphi(r)}^{\varphi(r)}\int_{0}^{r}\lambda_\text{T} p_\text{L}(x)x\mathrm{d}x\mathrm{d}\theta \nonumber\\&+ \int_{ \varphi(r)}^{2\pi-\varphi(r)}\int_{0}^{R(\theta)}\lambda_\text{T} p_\text{L}(x)x\mathrm{d}x\mathrm{d}\theta =2\lambda_\text{T}\varphi(r)\int_{0}^{r} xp_\text{L}(x)\mathrm{d}x\nonumber\hspace{0pt}\\&+\lambda_\text{T}\int_{\varphi(r)}^{2\pi- \varphi(r)}\int_{0}^{R(\theta)} xp_\text{L}(x) \mathrm{d}x \mathrm{d}\theta,\hspace{0pt}
\end{align}
where $\varphi (r) =  {\cos ^{ - 1}}\left( {\frac{{{r^2} + {d^2} - {D^2}}}{{2dr}}} \right)$ and $R(\theta)=\sqrt {{D^2}-{d^2}{{\sin }^2}\left( \theta  \right) }+d\cos \left( \theta  \right)$. Also, in Case 2, we define ${\cal H}_d(r) = \frac{1}{\lambda_\text{T}}{\Lambda}_\text{L}({\rm{\cal C}_r})$ for notational simplicity and also to make it independent of the deployment intensity $\lambda_\text{T}$.

Then, the cumulative distribution function (CDF) of ${\| \mathbf{x}_\text{L} \|}$ is
\begin{align}
&{\mathbb{P}}\left( \| \mathbf{x}_\text{L} \| \leq r\right) =1-{\mathbb{P}}\left( \| \mathbf{x}_\text{L} \| > r \right)=\nonumber\hspace{+0pt}\\ &\left\{ \begin{array}{*{20}{c}}
{\frac{1-\exp\left(-2 \pi \lambda_\text{T} \int_{0}^{r} x p_\text{L}(x) \mathrm{d}x\right)}{1-\exp(-\lambda_\text{T} {\cal H}_d)} \hspace{10pt} 0\leq r<D-d,}\\
{\frac{1- \exp\left(- \lambda_\text{T} {\cal H}_d(r)\right)}{{1-\exp(-\lambda_\text{T} {\cal H}_d)}} \hspace{+22pt} D-d \leq r<D+d,}\\
{1 \hspace{+123pt} r> D+d,}
\end{array}\right. \hspace{0pt}
\end{align}
where ${\cal H}_d$ is defined as
\begin{align}
{\cal H}_d =  \frac{1}{\lambda_\text{T}}{\Lambda}_\text{L}({\rm{\cal A}}) ={\cal H}_d(D+d) = \int_{ 0}^{2\pi}\int_{0}^{R(\theta)} xp_\text{L}(x) \mathrm{d}x \mathrm{d}\theta.
\end{align}

The probability density function (PDF) can be obtained by taking derivation from the CDF, which leads to
\begin{align}
&{f_{\text{T},\text{L}}^d}\left( r\right) = \nonumber\\&\left\{ {\begin{array}{*{20}{c}}
{\frac{2\pi \lambda_\text{T} r p_\text{L}(r)\exp\left(-2 \pi \lambda_\text{T} \int_{0}^{r} xp_\text{L}(x) \mathrm{d}x\right)}{1-\exp(-\lambda_\text{T} {\cal H}_d)} \hspace{+5pt} 0\leq r<D-d,}\\
{ \frac{\lambda_\text{T} \frac{{\partial {\cal H}_d(r) }}{{\partial r}} \exp\left(- \lambda_\text{T} {\cal H}_d(r)\right)}{1-\exp(-\lambda_\text{T} {\cal H}_d)} \hspace{+23pt} D-d \leq r<D+d,}\\
{0 \hspace{+148pt} r\ge D+d.}
\end{array}} \right. 
\end{align}

Following a similar approach for the distance distribution of ${\| \mathbf{x}_\text{L} \|}$, the CDF of the distance of the reference receiver to its closest NLOS transmitter, i.e., ${\| \mathbf{x}_\text{N} \|}$, is given by
\begin{align}
&{\mathbb{P}}\left( {{\| \mathbf{x}_\text{N} \|} \leq r} \right) =\nonumber\\ &\left\{ {\begin{array}{*{20}{c}}
{\frac{1-\exp\left(-2 \pi \lambda_\text{T} \int_{0}^{r} x(1-p_\text{L}(x)) \mathrm{d}x\right)}{1-\exp(-\lambda_\text{T} {\cal G}_d)} \hspace{+10pt} 0\leq r<D-d,}\\
{\frac{1- \exp\left(- \lambda_\text{T} {\cal G}_d(r)\right)}{1-\exp(-\lambda_\text{T} {\cal G}_d)} \hspace{41pt} D-d \leq r<D+d,}\\
{1 \hspace{+139pt} r> D+d,}
\end{array}} \right. 
\end{align}
where ${\cal G}_d(r)$ is defined as
\begin{align}
{\cal G}_d(r) &= 2\varphi(r)\int_{0}^{r} x(1-p_\text{L}(x))\mathrm{d}x\nonumber\\&+\int_{ \varphi(r)}^{2\pi-\varphi(r)}\int_{0}^{R(\theta)} x(1-p_\text{L}(x)) \mathrm{d}x \mathrm{d}\theta,
\end{align}
and ${\cal G}_d= \int_{ 0}^{2\pi}\int_{0}^{R(\theta)} x(1-p_\text{L}(x)) \mathrm{d}x \mathrm{d}\theta$. Then, the PDF of ${\| \mathbf{x}_\text{N} \|}$ is found as
\begin{align}
&{f_{\text{T},\text{N}}^d}\left( r\right) =\nonumber\\& \left\{ {\begin{array}{*{20}{c}}
{\frac{2\pi \lambda_\text{T} r (1-p_\text{L}(r))\exp\left(-2 \pi \lambda_\text{T} \int_{0}^{r} x(1-p_\text{L}(x)) \mathrm{d}x\right)}{1-\exp(-\lambda_\text{T} {\cal G}_d)} \hspace{+5pt} 0\leq r<D-d,}\\
{\frac{\lambda_\text{T} \frac{{\partial {\cal G}_d(r) }}{{\partial r}} \exp\left(- \lambda_\text{T} {\cal G}_d(r)\right)}{1-\exp(-\lambda_\text{T} {\cal G}_d)} \hspace{+60pt} D-d \leq r<D+d,}\\
{0 \hspace{+182pt} r > D+d.}
\end{array}} \right.
\end{align} 
Using the PDF and the CDF of the distances of the reference receiver to its closest LOS and NLOS transmitters, the association probabilities of the receiver in connection to an LOS and NLOS transmitter are given in the following theorems.
\begin{theorem}
The association probability that the reference receiver is served by an LOS transmitter in the case $(ِD-d)^{\frac{\alpha_\text{N}}{\alpha_\text{L}}} > D+d$ is
\begin{align}
\hspace{-10pt}{\cal A}_{\text{T},\text{L}}^1 (d) &= \int_{0}^{D-d} 2\pi \lambda_\text{T} r p_\text{L}(r)\exp\Biggl(-2 \pi \lambda_\text{T}\times \nonumber\hspace{0pt}\\&\Biggl(\int_{0}^{r^{\frac{\alpha_\text{L}}{\alpha_\text{N}}}} x(1-p_\text{L}(x)) \mathrm{d}x+ \int_{0}^{r} xp_\text{L}(x) \mathrm{d}x\Biggr)\Biggr) \mathrm{d}r\nonumber\\&+\int_{ِD-d}^{D+d}\lambda_\text{T} \frac{{\partial {\cal H}_d(r) }}{{\partial r}}\exp\Biggl(-2 \pi \lambda_\text{T} \times\nonumber\\&\Biggl(\int_{0}^{r^{\frac{\alpha_\text{L}}{\alpha_\text{N}}}} x(1-p_\text{L}(x)) \mathrm{d}x+\frac{1}{2\pi} {\cal H}_d(r)\Biggr)\Biggr)\mathrm{d}r,\hspace{0pt}
\end{align}
and in the case $(ِD-d)^{\frac{\alpha_\text{N}}{\alpha_\text{L}}} < D+d$ is given by
\begin{align}
{\cal A}_{\text{T},\text{L}}^2 (d) &=  \int_{0}^{D-d} 2\pi \lambda_\text{T} r p_\text{L}(r)\exp\Biggl(-2 \pi \lambda_\text{T}\times\nonumber\hspace{50pt}\\
 &\Biggl(\int_{0}^{r^{\frac{\alpha_\text{L}}{\alpha_\text{N}}}} x(1-p_\text{L}(x)) \mathrm{d}x+ \int_{0}^{r} xp_\text{L}(x) \mathrm{d}x\Biggr)\Biggr) \mathrm{d}r\nonumber\hspace{0pt}\\
&+\int_{ِD-d}^{(ِD-d)^{\frac{\alpha_\text{N}}{\alpha_\text{L}}}}\lambda_\text{T} \frac{{\partial {\cal H}_d(r) }}{{\partial r}}\exp\Biggl(-2 \pi \lambda_\text{T} \times \nonumber\\
 &\Biggl(\int_{0}^{r^{\frac{\alpha_\text{L}}{\alpha_\text{N}}}} x(1-p_\text{L}(x)) \mathrm{d}x+\frac{1}{2\pi} {\cal H}_d(r)\Biggr)\Biggr)\mathrm{d}r\nonumber\hspace{0pt}\\
&\hspace{-40pt}+\int_{(ِD-d)^{\frac{\alpha_\text{N}}{\alpha_\text{L}}}}^{D+d} \lambda_\text{T} \frac{{\partial {\cal H}_d(r) }}{{\partial r}} \exp\biggl(- \lambda_\text{T} \biggl({\cal G}_d\left(r^{\frac{\alpha_\text{L}}{\alpha_\text{N}}}\right) + {\cal H}_d(r)\biggr)\biggr)\mathrm{d}r.\hspace{0pt}
\end{align}
\end{theorem}
\begin{IEEEproof}
See Appendix A.
\end{IEEEproof}
\begin{theorem}
The association probability that the reference receiver is served by an NLOS transmitter in the case $(ِD-d)^{\frac{\alpha_\text{N}}{\alpha_\text{L}}} > D+d$ is
\begin{align}
\hspace{0pt}{\cal A}_{\text{T},\text{N}}^1 (d) &=\int_{0}^{(ِD-d)^{\frac{\alpha_\text{L}}{\alpha_\text{N}}}} 2\pi \lambda_\text{T} r (1-p_\text{L}(r))\exp\Biggl(-2 \pi \lambda_\text{T} \times\nonumber\\&\Biggl(\int_{0}^{r^{\frac{\alpha_\text{N}}{\alpha_\text{L}}}} xp_\text{L}(x) \mathrm{d}x+ \int_{0}^{r} x(1-p_\text{L}(x)) \mathrm{d}x\Biggr)\Biggr) \mathrm{d}r\nonumber\\&+\int_{(ِD-d)^{\frac{\alpha_\text{L}}{\alpha_\text{N}}}}^{(ِD+d)^{\frac{\alpha_\text{L}}{\alpha_\text{N}}}}2\pi \lambda_\text{T} r (1-p_\text{L}(r))\exp\Biggl(-2 \pi \lambda_\text{T}\times\nonumber\\
&\Biggl(\frac{1}{2\pi}{\cal H}_d\left(r^\frac{\alpha_\text{N}}{\alpha_\text{L}}\right)+ \int_{0}^{r} x(1-p_\text{L}(x)) \mathrm{d}x\Biggr)\Biggr) \mathrm{d}r \nonumber\hspace{0pt}\\&+\int_{(ِD+d)^{\frac{\alpha_\text{L}}{\alpha_\text{N}}}}^{D-d}2\pi \lambda_\text{T} r (1-p_\text{L}(r))\times\nonumber\\&\exp\Biggl(-2 \pi \lambda_\text{T} \Biggl(\frac{1}{2\pi}{\cal H}_d+ \int_{0}^{r} x(1-p_\text{L}(x)) \mathrm{d}x\Biggr)\Biggr) \mathrm{d}r \nonumber\hspace{0pt}\\&+\int_{ِD-d}^{D+d} \lambda_\text{T} \frac{{\partial {\cal G}_d(r) }}{{\partial r}} \exp\bigl(- \lambda_\text{T} \bigl({\cal H}_d+ {\cal G}_d(r)\bigr)\bigr)\mathrm{d}r,\hspace{0pt}
\end{align}
and in the case $(ِD-d)^{\frac{\alpha_\text{N}}{\alpha_\text{L}}} < D+d$ is given by
\begin{align}
\hspace{0pt}{\cal A}_{\text{T},\text{N}}^2 (d) = \int_{0}^{(ِD-d)^{\frac{\alpha_\text{L}}{\alpha_\text{N}}}} 2\pi \lambda_\text{T} r (1-p_\text{L}(r))\exp\biggl(-2 \pi \lambda_\text{T} \times \nonumber\\\biggl(\int_{0}^{r^{\frac{\alpha_\text{N}}{\alpha_\text{L}}}} xp_\text{L}(x) \mathrm{d}x+ \int_{0}^{r} x(1-p_\text{L}(x)) \mathrm{d}x\biggr)\biggr) \mathrm{d}r\nonumber\\+\int_{(ِD-d)^{\frac{\alpha_\text{L}}{\alpha_\text{N}}}}^{D-d}2\pi \lambda_\text{T} r (1-p_\text{L}(r))\exp\biggl(-2 \pi \lambda_\text{T}\times \nonumber\hspace{5pt}\\\left(\frac{1}{2\pi}{\cal H}_d\left(r^\frac{\alpha_\text{N}}{\alpha_\text{L}}\right)+ \int_{0}^{r} x(1-p_\text{L}(x)) \mathrm{d}x\right)\biggr) \mathrm{d}r\nonumber\hspace{10pt}\\+\int_{D-d}^{{(ِD+d)^{\frac{\alpha_\text{L}}{\alpha_\text{N}}}}} \lambda_\text{T} \frac{{\partial {\cal G}_d(r) }}{{\partial r}} \exp\left(- \lambda_\text{T} \left({\cal H}_d\left(r^{\frac{\alpha_\text{N}}{\alpha_\text{L}}}\right) + {\cal G}_d(r)\right)\right)\mathrm{d}r\nonumber\hspace{0pt}
\end{align}
\vspace{-10pt}
\begin{align}
+\int_{{(ِD+d)^{\frac{\alpha_\text{L}}{\alpha_\text{N}}}}}^{D+d} \lambda_\text{T} \frac{{\partial {\cal G}_d(r) }}{{\partial r}} \exp\left(- \lambda_\text{T} \left({\cal H}_d+ {\cal G}_d(r)\right)\right)\mathrm{d}r.\hspace{0pt}
\end{align}
\end{theorem}
\begin{IEEEproof}
The proof follows the same approach as in Appendix A, except that (8) and (13) are used instead of (11) and (10), respectively. Thus, due to space limit, the proof is omitted.
\end{IEEEproof}
Using the association probabilities, the distance distributions of the serving transmitter conditioned on the association of the reference receiver to an LOS and NLOS transmitter are presented in the following theorems.
\begin{theorem}
If a receiver is served by an LOS transmitter, the PDF of the distance to its serving transmitter in the case $(ِD-d)^{\frac{\alpha_\text{N}}{\alpha_\text{L}}} > D+d$ is
\begin{align}
&{{\hat f}_{\text{T},\text{L}}^{d,1}}(r) =\nonumber\\& \left\{\begin{matrix}
 \frac{2\pi \lambda_\text{T}rp_\text{L}(r)}{{\cal A}_{\text{T},\text{L}}^1(d)}\exp\biggl(-2 \pi \lambda_\text{T} \biggl(\int_{0}^{r^{\frac{\alpha_\text{L}}{\alpha_\text{N}}}} x(1-p_\text{L}(x)) \mathrm{d}x\hspace{20pt}\\+ \int_{0}^{r} xp_\text{L}(x) \mathrm{d}x\biggr)\biggr) \hspace{80pt}& \hspace{-130pt}\text{if}  \hspace{+2pt} 0<r<D-d,\\ 
\frac{\lambda_\text{T}}{{\cal A}_{\text{T},\text{L}}^1(d)} \frac{{\partial {\cal H}_d(r) }}{{\partial r}}\exp\biggl(-2 \pi \lambda_\text{T}\biggl(\int_{0}^{r^{\frac{\alpha_\text{L}}{\alpha_\text{N}}}} x(1-p_\text{L}(x)) \mathrm{d}x\hspace{5pt}\\+\frac{1}{2\pi} {\cal H}_d(r)\biggr)\biggr)\hspace{95pt}&\hspace{-110pt} \text{if}  \hspace{+2pt} D-d<r<D+d,\\ 
0\hspace{180pt} & \hspace{-149pt}\text{if}  \hspace{+2pt} r>D+d,
 \end{matrix}\right.
\end{align}
and in the case $(ِD-d)^{\frac{\alpha_\text{N}}{\alpha_\text{L}}} < D+d$ is given by
\begin{align}
&{{\hat f}_{\text{T},\text{L}}^{d,2}}(r) =\nonumber\\& \left\{\begin{matrix}
 \frac{2\pi \lambda_\text{T}rp_\text{L}(r)}{{\cal A}_{\text{T},\text{L}}^2(d)}\exp\biggl(-2 \pi \lambda_\text{T} \biggl(\int_{0}^{r^{\frac{\alpha_\text{L}}{\alpha_\text{N}}}} x(1-p_\text{L}(x)) \mathrm{d}x\hspace{13pt}\\+ \int_{0}^{r} xp_\text{L}(x) \mathrm{d}x\biggr)\biggr)\hspace{+120pt} & \hspace{-127pt}\text{if}  \hspace{+2pt} 0<r<D-d,\\ 
\frac{\lambda_\text{T}}{{\cal A}_{\text{T},\text{L}}^2(d)} \frac{{\partial {\cal H}_d(r) }}{{\partial r}}\exp\biggl(-2 \pi \lambda_\text{T} \biggl(\int_{0}^{r^{\frac{\alpha_\text{L}}{\alpha_\text{N}}}} x(1-p_\text{L}(x)) \mathrm{d}x\\+\frac{1}{2\pi} {\cal H}_d(r)\biggr)\biggr)\hspace{141pt}&\hspace{-85pt} \text{if}  \hspace{+2pt} D-d<r<(ِD-d)^{\frac{\alpha_\text{N}}{\alpha_\text{L}}},\\
\frac{\lambda_\text{T}}{{\cal A}_{\text{T},\text{L}}^2(d)} \frac{{\partial {\cal H}_d(r) }}{{\partial r}} \times\hspace{140pt}\\\exp\left({- \lambda_\text{T} \left({\cal G}_d\left(r^{\frac{\alpha_\text{L}}{\alpha_\text{N}}}\right) + {\cal H}_d(r)\right)}\right)\hspace{70pt}&\hspace{-80pt} \text{if}  \hspace{+5pt} (ِD-d)^{\frac{\alpha_\text{N}}{\alpha_\text{L}}}<r<D+d,\\
0 \hspace{+185pt}&\hspace{-140pt} \text{if}  \hspace{+2pt} r>D+d.
 \end{matrix}\right.
\end{align}
\end{theorem}
\begin{IEEEproof}
See Appendix B.
\end{IEEEproof}
\begin{theorem}
If a receiver is served by an NLOS transmitter, the PDF of the distance to its serving transmitter in the case $(ِD-d)^{\frac{\alpha_\text{N}}{\alpha_\text{L}}} > D+d$ is
\begin{align}
&{{\hat f}_{\text{T},\text{N}}^{d,1}}(r) =\nonumber\\& \left\{\begin{matrix}
\frac{2\pi \lambda_\text{T} r (1-p_\text{L}(r))}{{\cal A}_{\text{T},\text{N}}^1(d)}\exp\biggl(-2 \pi \lambda_\text{T} \biggl(\int_{0}^{r^{\frac{\alpha_\text{N}}{\alpha_\text{L}}}} xp_\text{L}(x) \mathrm{d}x\\+ \int_{0}^{r} x(1-p_\text{L}(x)) \mathrm{d}x\biggr)\biggr)\hspace{80pt} &\hspace{-127pt} \text{if}  \hspace{+2pt} 0<r<{(ِD-d)^{\frac{\alpha_\text{L}}{\alpha_\text{N}}}},\\ 
\frac{2\pi \lambda_\text{T} r (1-p_\text{L}(r))}{{\cal A}_{\text{T},\text{N}}^1(d)}\exp\biggl(-2 \pi \lambda_\text{T} \left(\frac{1}{2\pi}{\cal H}_d\biggl(r^\frac{\alpha_\text{N}}{\alpha_\text{L}}\right)\hspace{10pt}\\+ \int_{0}^{r} x(1-p_\text{L}(x)) \mathrm{d}x\biggr)\biggr) \hspace{80pt}&\hspace{-90pt} \text{if}  \hspace{+2pt} {(ِD-d)^{\frac{\alpha_\text{L}}{\alpha_\text{N}}}}<r<{(ِD+d)^{\frac{\alpha_\text{L}}{\alpha_\text{N}}}},\\
\frac{2\pi \lambda_\text{T} r (1-p_\text{L}(r))}{{\cal A}_{\text{T},\text{N}}^1(d)}\exp\biggl(-2 \pi \lambda_\text{T} \biggl(\frac{1}{2\pi}{\cal H}_d\hspace{47pt}\\+ \int_{0}^{r} x(1-p_\text{L}(x)) \mathrm{d}x\biggr)\biggr)\hspace{80pt} &\hspace{-103pt}\text{if}  \hspace{+2pt} {(ِD+d)^{\frac{\alpha_\text{L}}{\alpha_\text{N}}}}<r<D-d,\\
\frac{\lambda_\text{T}}{{\cal A}_{\text{T},\text{N}}^1(d)} \frac{{\partial {\cal G}_d(r) }}{{\partial r}}\times \hspace{130pt}\\\exp\left({- \lambda_\text{T} \left({\cal H}_d+ {\cal G}_d(r)\right)}\right) \hspace{75pt}&\hspace{-122pt} \text{if}  \hspace{+2pt} D-d<r<D+d,\\
0 \hspace{170pt}&\hspace{-161pt}\text{if}  \hspace{+2pt} r>D+d,
 \end{matrix}\right.
\end{align}
and in the case $(ِD-d)^{\frac{\alpha_\text{N}}{\alpha_\text{L}}} < D+d$ is given by
\begin{align}
&{{\hat f}_{\text{T},\text{N}}^{d,2}}(r) =\nonumber\\&\hspace{-10pt} \left\{\begin{matrix}
\frac{2\pi \lambda_\text{T} r (1-p_\text{L}(r))}{{\cal A}_{\text{T},\text{N}}^2(d)}\exp\biggl(-2 \pi \lambda_\text{T} \biggl(\int_{0}^{r^{\frac{\alpha_\text{N}}{\alpha_\text{L}}}} xp_\text{L}(x) \mathrm{d}x\hspace{15pt}\\+ \int_{0}^{r} x(1-p_\text{L}(x)) \mathrm{d}x\biggr)\biggr) \hspace{90pt}& \hspace{-95pt}\text{if}  \hspace{+2pt} 0<r<{(ِD-d)^{\frac{\alpha_\text{L}}{\alpha_\text{N}}}},\\
\frac{2\pi \lambda_\text{T} r (1-p_\text{L}(r))}{{\cal A}_{\text{T},\text{N}}^2(d)}\exp\biggl(-2 \pi \lambda_\text{T} \biggl(\frac{1}{2\pi}{\cal H}_d\left(r^\frac{\alpha_\text{N}}{\alpha_\text{L}}\right)\hspace{30pt}\\+ \int_{0}^{r} x(1-p_\text{L}(x)) \mathrm{d}x\biggr)\biggr) \hspace{90pt}&\hspace{-75pt} \text{if}  \hspace{+2pt} {(ِD-d)^{\frac{\alpha_\text{L}}{\alpha_\text{N}}}}<r<D-d,\\
\frac{\lambda_\text{T}}{{\cal A}_{\text{T},\text{N}}^2(d)} \frac{{\partial {\cal G}_d(r) }}{{\partial r}}\times\hspace{140pt} \\\exp\left(- \lambda_\text{T} \left({\cal H}_d\left(r^{\frac{\alpha_\text{N}}{\alpha_\text{L}}}\right) + {\cal G}_d(r)\right)\right)\hspace{60pt}& \hspace{-73pt}\text{if}  \hspace{+2pt} D-d<r<{(ِD+d)^{\frac{\alpha_\text{L}}{\alpha_\text{N}}}},\\
\frac{\lambda_\text{T}}{{\cal A}_{\text{T},\text{N}}^2(d)} \frac{{\partial {\cal G}_d(r) }}{{\partial r}} \times \hspace{140pt}\\\exp\left({- \lambda_\text{T} \left({\cal H}_d+ {\cal G}_d(r)\right)}\right)\hspace{100pt}& \hspace{-73pt}\text{if}  \hspace{+2pt} {(ِD+d)^{\frac{\alpha_\text{L}}{\alpha_\text{N}}}}<r<D+d,\\
0 \hspace{180pt}&\hspace{-133pt} \text{if}  \hspace{+2pt} r>D+d.
 \end{matrix}\right.
\end{align}
\end{theorem}
\begin{IEEEproof}
The proof follows the same approach as in Appendix B, except that (8) and (13) are used instead of (11) and (10), respectively. Thus, due to space limit, the proof is omitted.
\end{IEEEproof}
\section{Coverage Probability and Ergodic Rate Analysis}
In this section, the distance distribution results and association probabilities in (14)-(21) are used to derive the coverage probability and the ergodic rate for the reference receiver. 

The coverage probability given the minimum required SINR $\beta$ can be computed as
\begin{align}
&{P}_\text{C}^i (d,\beta) =\nonumber\\& {\cal A}_{\text{T},\text{L}}^i(d) {P}_{\text{C},\text{L}}^i (d,\beta)+ {\cal A}_{\text{T},\text{N}}^i(d) {P}_{\text{C},\text{N}}^i (d,\beta),\ i =\left\{1,2\right\},
\end{align}
where $i=1$ denotes the case $(ِD-d)^{\frac{\alpha_\text{N}}{\alpha_\text{L}}} > D+d$ and $i=2$ is for $(ِD-d)^{\frac{\alpha_\text{N}}{\alpha_\text{L}}} < D+d$. In addition, ${\cal A}_{\text{T},\text{L}}^i(d)$ and ${\cal A}_{\text{T},\text{N}}^i(d)$ are the association probabilities derived in Theorems 1 and 2, respectively. Also, ${P}_{\text{C},\text{L}}^i(d,\beta)$ and ${P}_{\text{C},\text{N}}^i(d,\beta)$ are the conditional coverage probability given that the receiver is associated with an LOS and NLOS transmitter, respectively. Note that the coverage probability is zero when there is no transmitter inside $\cal A$, which happens with probability $1-{\cal A}_{\text{T},\text{L}}^i(d)-{\cal A}_{\text{T},\text{N}}^i(d)$.

In the case of the association to an LOS transmitter, i.e., $q=\text{L}$ in (3), the conditional coverage probability ${P}_{\text{C},\text{L}}^i(d,\beta)$ is found as 
\begin{align}
&{P}_{\text{C},\text{L}}^i (d,\beta)= \mathbb{P}\left(\frac{{{a_1}{h_\text{L}}{{r}^{ - {\alpha _\text{L}}}}}}{{{\sigma ^2} + {{\cal I}_\text{L}} + {{\cal I}_\text{N}}}}>\beta \right) \nonumber\\&= \int_{0}^{D+d}\mathbb{P}\left(\frac{{{a_1}{h_\text{L}}{{r}^{ - {\alpha _\text{L}}}}}}{{{\sigma ^2} + {{\cal I}_\text{L}} + {{\cal I}_\text{N}}}}>\beta\right) {\hat f}_{\text{T},\text{L}}^{d,i} (r) \mathrm{d}r,
\end{align}
where ${\hat f}_{\text{T},\text{L}}^{d,i}$ is given in Theorem 3, and the conditional coverage probability given a link distance $r$ is obtained as
\begin{align}
&\mathbb{P}\left(\frac{{{a_1}{h_\text{L}}{{r}^{ - {\alpha _\text{L}}}}}}{{{\sigma ^2} + {{\cal I}_\text{L}} + {{\cal I}_\text{N}}}}>\beta\right) = \mathbb{P}\left(h_\text{L}>\frac{\beta r^{\alpha_\text{L}}}{a_1}\left({\sigma ^2} + {{\cal I}_\text{L}} + {{\cal I}_\text{N}}\right)\right)\nonumber\\&
\stackrel{(a)}{\approx}\mathop \sum \limits_{n = 1}^{{v_\text{L}}} {\left( { - 1} \right)^{n + 1}} {{v_\text{L}}\choose{n}} \mathbb{E}\left\{ {{e^{ - \frac{ \eta_\text{L} \beta n r^{{\alpha}_\text{L}}}{a_1} \left( {\sigma ^2} + {{\cal I}_\text{L}} + {{\cal I}_\text{N}}  \right)}}} \right\}\nonumber\hspace{+0pt}\\
&\stackrel{(b)}{=}\mathop \sum \limits_{n = 1}^{{v_\text{L}}} {\left( { - 1} \right)^{n + 1}}{{v_\text{L}}\choose{n}} e^{ - \frac{\eta_\text{L} \beta n r^{{\alpha}_\text{L}}}{a_1} {\sigma ^2}}{\cal L}_\text{L|L}\left(\frac{\eta_\text{L}\beta n r^{{\alpha}_\text{L}}}{a_1}\mid r\right)\nonumber\\&\hspace{15pt}\times{\cal L}_\text{N|L}\left(\frac{\eta_\text{L}\beta n r^{{\alpha}_\text{L}}}{a_1}\mid r\right),
\end{align}
where $\eta_\text{L} ={v_\text{L}}{\left( {{v_\text{L}}!} \right)^{ - \frac{1}{{{v_\text{L}}}}}}$, $(a)$ follows from $h_\text{L}\sim {{\Gamma}}(v_\text{L},\frac{1}{v_\text{L}})$ and Alzer's Lemma [25], and $(b)$ comes from the independency of ${\Phi}_\text{L}$ and ${\Phi}_\text{N}$, the definitions of LT as ${\cal L}_\text{L|L} (s\mid r) = \mathbb{E}\left\{e^{-s \cal I_\text{L}}\mid q = \text{L}\hspace{2pt}\&\hspace{2pt} \|\mathbf{x}_\text{L}\|=r\right\}$ and ${\cal L}_\text{N|L} (s\mid r)= \mathbb{E}\left\{e^{-s \cal I_\text{N}}\mid q = \text{L}\hspace{2pt}\&\hspace{2pt} \|\mathbf{x}_\text{L}\|=r\right\}$.

We can obtain ${\cal L}_\text{L|L} (s\mid r)$ as
\begin{align}
{\cal L}_\text{L|L} (s\mid r)  =\nonumber\hspace{170pt}\\ \mathbb{E}\left\{\exp\left(-s\mathop \sum \nolimits_{\mathbf{y} \in {\Phi}_\text{L} \backslash \left\{\mathbf{x}_\text{L}\right\}} {G_\mathbf{y}}{h_\mathbf{y}}{{\| \mathbf{y} \|}^{-{\alpha}_\text{L}}}\right)\mid n_\text{L}\geq 1\right\}\nonumber\hspace{+0pt}\\=\mathbb{E}\Biggl\{\mathop \prod \limits_{\mathbf{y} \in {\Phi}_\text{L} \backslash \left\{\mathbf{x}_\text{L}\right\}} \exp\left(-s {G_\mathbf{y}}{h_\mathbf{y}}{{\| \mathbf{y} \|}^{-{\alpha}_\text{L}}}\right)\mid n_\text{L}\geq 1\Biggr\}\nonumber\hspace{+10pt}\\
\stackrel{(c)}{=}\exp\biggl(-\lambda_\text{T} \int_{{\cal A}\backslash \mathbf{b}(\mathbf{o},r)}^{} \biggl(1-\nonumber\hspace{107pt}\\\mathbb{E}_{h_\mathbf{y}, G_\mathbf{y}}\left\{\exp\left(-s {G_\mathbf{y}}{h_\mathbf{y}}{{\| \mathbf{y} \|}^{-{\alpha}_\text{L}}}\right)\right\}\biggr)p_\text{L}(\|\mathbf{y}\|)\mathrm{d}\mathbf{y}\biggr)\nonumber\hspace{6pt}\\\stackrel{(d)}{=}\exp\biggl(-\lambda_\text{T} \int_{{\cal A}\backslash \mathbf{b}(\mathbf{o},r)}^{} \biggl(1-\nonumber\hspace{106pt}\\\mathbb{E}_{G_\mathbf{y}}\left\{\left(1+\frac{s{G_\mathbf{y}}{{\| \mathbf{y} \|}^{-{\alpha}_\text{L}}}}{v_\text{L}}\right)^{-v_\text{L}}\right\}\biggr)p_\text{L}(\|\mathbf{y}\|)\mathrm{d}\mathbf{y}\biggr),\hspace{6pt}
\end{align}
where $(c)$ follows from the probability generating functional (PGFL) of the PPP [6, Thm. 4.9] and $(d)$ is obtained by the moment-generating function (MGF) of $h_\mathbf{y} \sim {{\Gamma}}(v_\text{L},\frac{1}{v_\text{L}})$. 

Defining $\measuredangle \mathbf{y}$ as the angle of the line crossing $\mathbf{y}$ and the origin, the transmitter at $\mathbf{y}$ has distance $\hat d (\mathbf{y}) =\sqrt {{d^2} + {{\| \mathbf{y} \|}^2} - 2d\| \mathbf{y} \|{\rm{cos}}\left( {\pi  - \measuredangle \mathbf{y}} \right)}$ to the center of $\cal A$ and is assumed to serve a receiver with distance $R_\text{T}(\mathbf{y})$ to $\mathbf{y}$. Therefore, according to the characterization of the random variable $G_\mathbf{y}$ in Cases 1-4, with defined $b_k$, for $k=1,\ldots,4$, in Appendix C, we have
\begin{eqnarray}
\mathbb{E}_{G_\mathbf{y}}\left\{\left(1+\frac{s {G_\mathbf{y}}{{\| \mathbf{y} \|}^{-{\alpha}_\text{L}}}}{v_\text{L}}\right)^{-v_\text{L}}\mid R_\text{T} (\mathbf{y})= y\right\} \nonumber\hspace{30pt}\\= \mathop \sum \limits_{k = 1}^4 {b_k}\left( \mathbf{y},y,r \right) \left(1+\frac{s a_k{{\| \mathbf{y} \|}^{-{\alpha}_\text{L}}}}{v_\text{L}}\right)^{-v_\text{L}}, 
\end{eqnarray}
and then by conditioning on the distance $R_\text{T}(\mathbf{y})$, the unconditional result required for (25) is found as
\begin{align}
&\mathbb{E}_{G_\mathbf{y}}\left\{\left(1+\frac{s {G_\mathbf{y}}{{\| \mathbf{y} \|}^{-{\alpha}_\text{L}}}}{v_\text{L}}\right)^{-v_\text{L}}\right\} =\nonumber\\&\hspace{-10pt}\int_{0}^{\infty} \mathbb{E}_{G_\mathbf{y}}\left\{\left(1+\frac{s {G_\mathbf{y}}{{\| \mathbf{y} \|}^{-{\alpha}_\text{L}}}}{v_\text{L}}\right)^{-v_\text{L}}\mid R_\text{T} (\mathbf{y})= y\right\} f_{R_\text{T}(\mathbf{y})}(y)\mathrm{d}y,
\end{align}
where $f_{R_\text{T}(\mathbf{y})}$ is the PDF of $R_\text{T}(\mathbf{y})$. Since the exact characterization of the correlations among the receivers and between the transmitters and their served receivers in the network is very complicated and for tractability and concreteness similar as in uplink use-case scenarios, e.g., [26]-[29], we assume that ${\hat \Phi}_\text{R}$ is an FHPPP. We also assume that the distances $R_\text{T}$ for different transmitters are independent, and $R_\text{T}(\mathbf{y})$ is equal to the distance of the transmitter at $\mathbf{y}$ to the selected receiver over ${\hat \Phi}_\text{R}$ based on the average received power selection strategy in Subsection II.C. In Section V, the accuracy of the assumptions are verified through comparing simulation and numerical results (Fig. 3). Then, by conditioning on the association of the transmitter to an LOS or NLOS receiver, we can characterize $f_{R_\text{T}(\mathbf{y})}$ as
\begin{align}
&{f_{R_\text{T}(\mathbf{y})}}\left( y \right) =\nonumber\\&\left\{\begin{matrix}
\hspace{-8pt}{\cal A}_{\text{R},\text{L}}^1(\hat d(\mathbf{y})) {{\hat f}_{\text{R},\text{L}}^{\hat d(\mathbf{y}),1}}(y)\\+ {\cal A}_{\text{R},\text{N}}^1(\hat d(\mathbf{y})) {{\hat f}_{\text{R},\text{N}}^{\hat d(\mathbf{y}),1}}(y)  & \text{if} \hspace{+2pt}(ِD-\hat d(\mathbf{y}))^{\frac{\alpha_\text{N}}{\alpha_\text{L}}} > D+\hat d(\mathbf{y}), \\ 
\hspace{-8pt}{\cal A}_{\text{R},\text{L}}^2(\hat d(\mathbf{y})) {{\hat f}_{\text{R},\text{L}}^{\hat d(\mathbf{y}),2}}(y)\\+ {\cal A}_{\text{R},\text{N}}^2(\hat d(\mathbf{y})) {{\hat f}_{\text{R},\text{N}}^{\hat d(\mathbf{y}),2}}(y) & \text{if} \hspace{+2pt}(ِD-\hat d(\mathbf{y}))^{\frac{\alpha_\text{N}}{\alpha_\text{L}}} < D+\hat d(\mathbf{y}),
\end{matrix}\right.
\end{align}
where ${\cal A}_{\text{R},\text{L}}^i(\hat d(\mathbf{y}))$, ${\cal A}_{\text{R},\text{N}}^i(\hat d(\mathbf{y}))$, ${{\hat f}_{\text{R},\text{L}}^{\hat d(\mathbf{y}),i}}(y)$, and ${{\hat f}_{\text{R},\text{N}}^{\hat d(\mathbf{y}),i}}(y)$ are defined the same as ${\cal A}_{\text{T},\text{L}}^i(\hat d(\mathbf{y}))$, ${\cal A}_{\text{T},\text{N}}^i(\hat d(\mathbf{y}))$, ${{\hat f}_{\text{T},\text{R}}^{\hat d(\mathbf{y}),i}}(y)$, and ${{\hat f}_{\text{T},\text{N}}^{\hat d(\mathbf{y}),i}}(y)$ in (14)-(21), respectively. 

Then, according to (25)-(28), we can compute ${\cal L}_\text{L|L} (s\mid r)$ as
\begin{align}
{\cal L}_\text{L|L} (s\mid r) = \exp\Biggl(-\lambda_\text{T} \int_{{\cal A}\backslash \mathbf{b}(\mathbf{o},r)}^{} \int_{0}^{D+d}\biggl(1-\nonumber\hspace{82pt}\\\mathop \sum \limits_{k = 1}^4 {b_k}\left(\mathbf{y}, y,r \right) \left(1+\frac{s a_k{{\| \mathbf{y} \|}^{-{\alpha}_\text{L}}}}{v_\text{L}}\right)^{-v_\text{L}}\biggr) f_{R_\text{T}(\mathbf{y})}(y)p_\text{L}(\|\mathbf{y}\|)\mathrm{d}y\mathrm{d}\mathbf{y}\Biggr).\hspace{0pt}
\end{align}
In order to convert (29) from Cartesian to polar coordinates, there are two cases:

\textbf{Case 1:}  If ${\cal A}\cap \mathbf{b}(\mathbf{o},r) = \mathbf{b}(\mathbf{o},r)$, i.e., $0\leq r<D-d$, then\footnote{$\mathbf{y}$ is a function of $x$ and $\theta$ in polar coordinates.}
\begin{align}
{\cal L}_\text{L|L} (s\mid r) = \exp\Biggl(-\lambda_\text{T} \int_{ 0}^{2\pi}\int_{r}^{R(\theta)} \int_{0}^{D+d}\biggl(1-\hspace{20pt}\nonumber\\\hspace{0pt}\mathop \sum \limits_{k = 1}^4 {b_k}\left(x, \theta,y,r \right) \left(1+\frac{s a_k{x^{-{\alpha}_\text{L}}}}{v_\text{L}}\right)^{-v_\text{L}}\biggr)\nonumber\\\times f_{R_\text{T}(x,\theta)}(y)p_\text{L}(x)x\mathrm{d}y\mathrm{d}{x}\mathrm{d}\theta \Biggr).\hspace{30pt}
\end{align}

\textbf{Case 2:} If ${\cal A}\cap \mathbf{b}(\mathbf{o},r) \ne \mathbf{b}(\mathbf{o},r)$, i.e., $D-d \leq r<D+d$, then
\begin{align}
{\cal L}_\text{L|L} (s\mid r) = \exp\Biggl(-\lambda_\text{T} \int_{ -\varphi(r)}^{\varphi(r)}\int_{r}^{R(\theta)} \int_{0}^{D+d}\biggl(1-\nonumber\hspace{15pt}\\\mathop \sum \limits_{k = 1}^4 {b_k}\left(x, \theta, y,r \right) \left(1+\frac{s a_k{x^{-{\alpha}_\text{L}}}}{v_\text{L}}\right)^{-v_\text{L}}\biggr)\nonumber\\\times f_{R_\text{T}(x,\theta)}(y)p_\text{L}(x)x\mathrm{d}y\mathrm{d}{x}\mathrm{d}\theta \Biggr).\hspace{30pt}
\end{align}
Following a similar approach as for ${\cal L}_\text{L|L} (s\mid r)$, we can obtain ${\cal L}_\text{N|L} (s\mid r)$ as
\begin{align}
&{\cal L}_\text{N|L} (s\mid r) =\nonumber\hspace{0pt}\\
 &\left\{\begin{matrix}
\exp\biggl(-\lambda_\text{T} \int_{ 0}^{2\pi}\int_{r^{\frac{\alpha_\text{L}}{\alpha_\text{N}}}}^{R(\theta)} \int_{0}^{D+d}\hspace{+75pt}\\\hspace{-2pt}\biggl\{\left(1-\mathop \sum \limits_{k = 1}^4 {b_k}\left( x,\theta, y,r \right) \left(1+\frac{s a_k{x^{-{\alpha}_\text{N}}}}{{v_\text{N}}}\right)^{-v_\text{N}}\right)\\ \hspace{-20pt}\times f_{R_\text{T}(x,\theta)}(y)(1-p_\text{L}(x))x\mathrm{d}y\mathrm{d}{x}\mathrm{d}\theta\biggr\} \biggr)  & \hspace{-58pt}\text{if} \hspace{+2pt}0<r<D-d, \\ 
\exp\biggl(-\lambda_\text{T} \int_{ -\varphi(r)}^{\varphi(r)}\int_{r^{\frac{\alpha_\text{L}}{\alpha_\text{N}}}}^{R(\theta)} \int_{0}^{D+d}\hspace{+65pt}\\\hspace{-2pt}\biggl\{\left(1-\mathop \sum \limits_{k = 1}^4 {b_k}\left(x, \theta, y,r \right) \left(1+\frac{s a_k{x^{-{\alpha}_\text{N}}}}{v_\text{N}}\right)^{-v_\text{N}}\right)\\ \hspace{-20pt}\times f_{R_\text{T}(x,\theta)}(y)(1-p_\text{L}(x))x\mathrm{d}y\mathrm{d}{x}\mathrm{d}\theta\biggr\} \biggr)  & \hspace{-37pt}\text{if} \hspace{+2pt}D-d<r<D+d.
\end{matrix}\right.
\end{align}
Here, note that interfering transmitters are outside of $\mathbf{b}(\mathbf{o},{r^{\frac{\alpha_\text{L}}{\alpha_\text{N}}}})$ since the serving transmitter is an LOS transmitter with distance $r$ to the origin. 

Following a similar approach as for ${P}_{\text{C},\text{L}}^i(d,\beta)$, we can also obtain ${P}_{\text{C},\text{N}}^i(d,\beta)$ as 
\begin{eqnarray}
{P}_{\text{C},\text{N}}^i(d,\beta) {\approx}\mathop \sum \limits_{n = 1}^{{v_\text{N}}} {\left( { - 1} \right)^{n + 1}}{{v_\text{N}}\choose{n}}\int_{0}^{D+d}e^{ - \frac{\eta_\text{N} \beta n r^{{\alpha}_\text{N}}}{a_1} {\sigma ^2}}\times\nonumber\\{\cal L}_\text{L|N}\left(\frac{\eta_\text{N}\beta n r^{{\alpha}_\text{N}}}{a_1}\mid r\right){\cal L}_\text{N|N}\left(\frac{\eta_\text{N}\beta n r^{{\alpha}_\text{N}}}{a_1}\mid r\right){\hat f}_{\text{T},\text{N}}^{d,i} (r) \mathrm{d}r,
\end{eqnarray}
where $\eta_\text{N} ={v_\text{N}}{\left( {{v_\text{N}}!} \right)^{ - \frac{1}{{{v_\text{N}}}}}}$ and ${\hat f}_{\text{T},\text{N}}^{d,i}$ is given in Theorem 4. Also, ${\cal L}_\text{L|N} (s\mid r) =\mathbb{E}\left\{e^{-s \cal I_\text{L}}\mid q = \text{N}\hspace{2pt}\&\hspace{2pt} \|\mathbf{x}_\text{N}\|=r\right\}$ and ${\cal L}_\text{N|N} (s\mid r) = \mathbb{E}\left\{e^{-s \cal I_\text{N}}\mid q = \text{N}\hspace{2pt}\&\hspace{2pt} \|\mathbf{x}_\text{N}\|=r\right\}$, which are given by
\begin{align}
&{\cal L}_\text{L|N} (s\mid r) =\nonumber\\& \left\{\begin{matrix}
\exp\biggl(-\lambda_\text{T} \int_{ 0}^{2\pi}\int_{r^\frac{\alpha_\text{N}}{\alpha_\text{L}}}^{R(\theta)} \int_{0}^{D+d}\hspace{+91pt}\\\biggl\{\left(1-\mathop \sum \limits_{k = 1}^4 {b_k}\left( x, \theta, y,r \right) \left(1+\frac{s a_k{x^{-{\alpha}_\text{L}}}}{v_\text{L}}\right)^{-v_\text{L}}\right)\hspace{15pt}\\ \times f_{R_\text{T}(x,\theta)}(y)p_\text{L}(x)x\mathrm{d}y\mathrm{d}{x}\mathrm{d}\theta \biggr\}\biggr)\hspace{54pt}  & \hspace{-77pt}\text{if} \hspace{+2pt}0<r<D-d, \\ 
\exp\biggl(-\lambda_\text{T} \int_{ -\varphi(r)}^{\varphi(r)}\int_{r^\frac{\alpha_\text{N}}{\alpha_\text{L}}}^{R(\theta)} \int_{0}^{D+d}\hspace{+79pt}\\\biggl\{\left(1-\mathop \sum \limits_{k = 1}^4 {b_k}\left(x, \theta, y,r \right) \left(1+\frac{s a_k{x^{-{\alpha}_\text{L}}}}{v_\text{L}}\right)^{-v_\text{L}}\right)\hspace{15pt}\\ \times f_{R_\text{T}(x,\theta)}(y)p_\text{L}(x)x\mathrm{d}y\mathrm{d}{x}\mathrm{d}\theta \biggr\} \biggr)\hspace{54pt}  & \hspace{-57pt}\text{if} \hspace{+2pt}D-d<r<D+d,
\end{matrix}\right.
\end{align}
\vspace{-15pt}
\begin{align}
&{\cal L}_\text{N|N} (s\mid r) =\nonumber\\ &\left\{\begin{matrix}
\exp\biggl(-\lambda_\text{T} \int_{ 0}^{2\pi}\int_{r}^{R(\theta)} \int_{0}^{D+d}\hspace{+96pt}\\\biggl\{\left(1-\mathop \sum \limits_{k = 1}^4 {b_k}\left( x, \theta, y,r \right) \left(1+\frac{s a_k{x^{-{\alpha}_\text{N}}}}{v_\text{N}}\right)^{-v_\text{N}}\right)\hspace{20pt}\\ \times f_{R_\text{T}(x,\theta)}(y)(1-p_\text{L}(x))x\mathrm{d}y\mathrm{d}{x}\mathrm{d}\theta \biggr\} \biggr) \hspace{48pt} &\hspace{-80pt} \text{if} \hspace{+2pt}0<r<D-d, \\ 
\exp\biggl(-\lambda_\text{T} \int_{ -\varphi(r)}^{\varphi(r)}\int_{r}^{R(\theta)} \int_{0}^{D+d}\hspace{+85pt}\\\biggl\{\left(1-\mathop \sum \limits_{k = 1}^4 {b_k}\left(x, \theta, y,r \right) \left(1+\frac{s a_k{x^{-{\alpha}_\text{N}}}}{v_\text{N}}\right)^{-v_\text{N}}\right)\hspace{20pt}\\ \times f_{R_\text{T}(x,\theta)}(y)(1-p_\text{L}(x))x\mathrm{d}y\mathrm{d}{x}\mathrm{d}\theta \biggr\} \biggr) \hspace{48pt}&\hspace{-62pt} \text{if} \hspace{+2pt}D-d<r<D+d.
\end{matrix}\right.
\end{align}

While the integrals can not be reduced to closed-form, it is easy to evaluate them numerically. 

For a receiver located at the center of $\cal A$, i.e., $d = 0$, the coverage probability is simplified since the results are independent of the angle of the line crossing each transmitter to the origin, which is due to the symmetry of the spatial model for $d=0$. Also, note that in the special case of infinite mmWave wireless networks, i.e., $D \to \infty$, the coverage probability analysis simplifies to the result in [9, Thm. 1]. The coverage probability for $d = 0$ (or $D \to \infty$) is not a lower or upper bound. This is because there is a tradeoff as these specific cases have two opposing effects on the coverage probability: i) distances (or the number) of both LOS and NLOS interfering transmitters decrease (or increases), which increases the interference power, and ii) the distance of the serving LOS or NLOS transmitter decreases, which increases the desired signal power. Also, as another effect for $d=0$, the transmitters are more likely to be LOS rather than being NLOS, which increases both the interference power and the desired signal power.

A lower/upper bound on the coverage probability in (22) can be obtained when we assume that all transmitters interfere on the reference receiver with their main/side antenna beams, i.e., $\theta_\text{T} = 2\pi$. Therefore, according to Appendix C, letting $j=1$ for the lower bound and $j=2$ for the upper bound, we replace $\mathbb{E}_{G_\mathbf{y}}\left\{\left(1+\frac{s {G_\mathbf{y}}{{\| \mathbf{y} \|}^{-{\alpha}_q}}}{v_q}\right)^{-v_q}\right\} =$
\begin{align}
\left\{\begin{matrix}
\frac{\theta_\text{R}}{2 \pi} \left(1+\frac{s a_{j}{{\| \mathbf{y} \|}^{-{\alpha}_q}}}{v_q}\right)^{-v_q}+\hspace{95pt}\\\left(1-\frac{\theta_\text{R}}{2 \pi}\right) \left(1+\frac{s a_{j+1}{{\| \mathbf{y} \|}^{-{\alpha}_q}}}{v_q}\right)^{-v_q}\hspace{75pt}  &\hspace{-100pt} \text{if} \hspace{+2pt}0<r<D-d, \\ 
d(\mathbf{y},r) \left(1+\frac{s a_{j}{{\| \mathbf{y} \|}^{-{\alpha}_q}}}{v_q}\right)^{-v_q}+\hspace{80pt}\\\left(1-d(\mathbf{y},r)\right) \left(1+\frac{s a_{j+1}{{\| \mathbf{y} \|}^{-{\alpha}_q}}}{v_q}\right)^{-v_q}\hspace{60pt}  &\hspace{-80pt} \text{if} \hspace{+2pt}D-d<r<D+d,
\end{matrix}\right.
\end{align}
for $q = \left\{\text{L},\text{N}\right\}$ instead of (27) and its equivalent NLOS expression into ${\cal L}_\text{L|L}$, ${\cal L}_\text{L|N}$, ${\cal L}_\text{N|L}$, and ${\cal L}_\text{N|N}$, where $d(\mathbf{y},r) =\max\left\{\frac{\min\left\{\hat \phi_\text{R} (\mathbf{y})+\frac{\theta_\text{R}}{2},\varphi (r)\right\}-\max\left\{\hat \phi_\text{R} (\mathbf{y})-\frac{\theta_\text{R}}{2},-\varphi (r)\right\}}{2\varphi (r)},0\right\}$. Also, $\hat \phi_\text{R} (\mathbf{y}) = \cos^{-1}\left(\frac{d^2+\|\mathbf{y}\|^2-{\hat d} (\mathbf{y})^2}{2d\|\mathbf{y}\|}\right)$.

The concluded lower and upper bounded coverage probabilities are much easier than the coverage probability in (22) to numerically evaluate since the bounds do not depend on the distance of an interfering transmitter to its served receiver in computations. 
 
Finally, the ergodic rate of the reference receiver in bandwidth $W$, defined as $\tau = W\mathbb{E}\left\{\text{log}(1+\text{SINR})\right\}$, can be obtained from the coverage probability as, e.g., [5, Thm. 3]
\begin{align}
\tau^i (d) &= \int_{0}^{\infty} W\mathbb{P}(\text{log}\left(1+\text{SINR})>t\mid i , d\right)\mathrm{d}t \nonumber\\
&= \frac{W}{\text{ln}2}\int_{0}^{\infty}\frac{{P}_\text{C}^i (d,t)}{t+1} \mathrm{d}t, \ i=\left\{1,2\right\}.
\end{align}
 
\section{results and discussion}
In this section, we consider a scenario of finite mmWave wireless networks in which the transmitters and receivers are distributed according to FHPPPs with intensity $\lambda_\text{T} = 0.004 \hspace{+2pt}\text{m}^{-2}$ and $\lambda_\text{R} = 0.04 \hspace{+2pt}\text{m}^{-2}$ in a disk with radius $D = 50 \hspace{+2pt}\text{m}$, respectively, and evaluate the coverage probability and the ergodic rate results derived in Section IV. We also provide Monte Carlo simulations to validate the accuracy of the results. While we presented the analytical results for a general function $p_\text{L}(r)$, here we focus on $p_\text{L}(r) = e^{-\mu r}$ as in the 3GPP blockage model [8], where the blockage exponent $\mu$ is a constant that depends on the geometry and density of the blockage process. Also, we consider uniform planar square antennas at the transmitters and the receivers that have the following equations between their main-lobe gain $M_q$ and side-lobe gain $m_q$ with their beamwidth $\theta_q$ [22]:
\begin{eqnarray}
M_q = \frac{3}{{\theta}_q^2}, \ m_q = \frac{\sqrt{3}{\theta}_q-\frac{3\sqrt{3}}{2\pi} \sin\left(\frac{{\theta}_q}{2}\right)}{\sqrt{3}{\theta}_q-\frac{\sqrt{3}}{2\pi}{\theta}_q^2 \sin\left(\frac{{\theta}_q}{2}\right)}, \ q = \left\{\text{T},\text{R}\right\}.
\end{eqnarray}
We further consider that $\theta_\text{T} = \theta_\text{R} = \theta$. The values of the parameters in Table I are used, unless otherwise stated. We further define the normalized (relative) distance $\delta = \frac{d}{D}$.

In Fig. 3, the analytical results and Monte Carlo simulations for the coverage probability are shown as a function of the minimum required SINR $\beta$, considering $\delta = \frac{1}{5}$, $\frac{3}{5}$, and $\frac{4}{5}$. It is observed that the analytical results tightly mimic the exact Monte Carlo results for different distances of the reference receiver from the center of the disk. Thus, the assumptions in Section IV can well be applied for the performance analysis of finite mmWave networks.

In the following, we study the impact of the distance of the receiver from the center of the disk, the beamwidth, and the blockage exponent on the coverage probability and the ergodic rate. We also investigate the tightness of the lower and upper bounds derived in Section IV.
\begin{table}[t]
\caption {Parameter Values} 
\vspace{-10pt}
\begin{center}
\resizebox{3.7cm}{!} {
    \begin{tabular}{| l | l |}
  
   \hline
    \hline
    \textbf{System Parameter} &{\textbf{Value}} \\ \hline
    $\lambda_\text{T}$ & 0.004 $\text{m}^{-2}$\\ \hline
    $\lambda_\text{R}$ & 0.04 $\text{m}^{-2}$\\ \hline
    $\sigma^2$ & -30 dB \\ \hline
    $(\theta_\text{T}, \theta_\text{R})$ & (36$^\circ$, 36$^\circ$) \\ \hline
    $D$ & 50 m \\ \hline
    $\mu$ & $\frac{1}{15}$ $\text{m}^{-1}$\\ \hline
    $(\alpha_\text{L}, \alpha_\text{N})$ & (2, 4) \\ \hline
    $(v_\text{L}, v_\text{N})$ & (3, 2) \\ \hline
    $W$ & 200 MHz \\ \hline
   \hline
    \end{tabular}}
 
\end{center}
\vspace{-15pt}
\end{table}
\begin{figure}[tb!]
\centering
\includegraphics[width =3.4in]{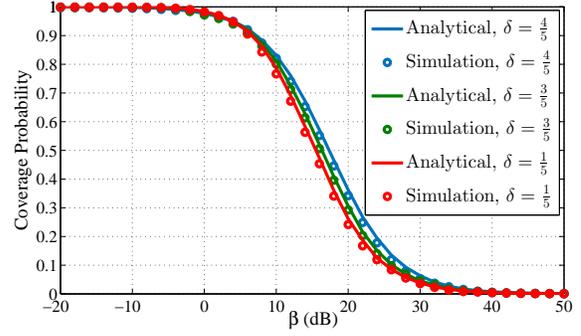}
\vspace{-5pt}
\caption{Coverage probability as a function of the SINR threshold $\beta$ for analytical and simulation results.}
\vspace{-5pt}
\end{figure}
\begin{figure}[t!]
\centering
\center
\vspace{-1ex}
\includegraphics[width =3.4in]{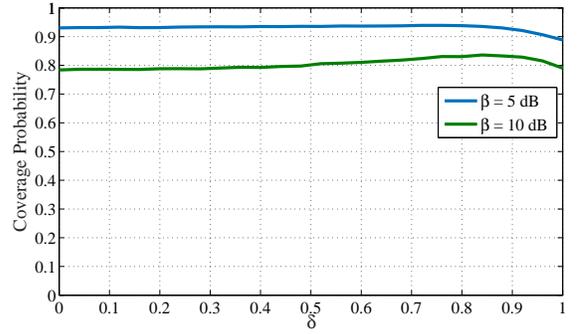}
\vspace{-5pt}
\caption{Coverage probability as a function of normalized distance $\delta$.}
\vspace{-5pt}
\end{figure}

\begin{figure}[t!]
\centering
\center
\vspace{-1ex}
\includegraphics[width =3.4in]{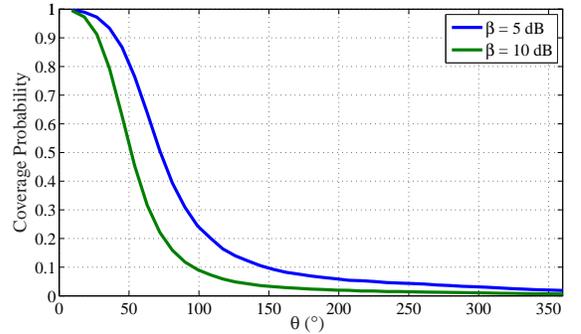}
\vspace{-5pt}
\caption{Coverage probability as a function of beamwidth $\theta$ with $\delta = \frac{2}{5}$.}
\vspace{-5pt}
\end{figure}

\begin{figure}[t!]
\centering
\center
\vspace{-1ex}
\includegraphics[width =3.4in]{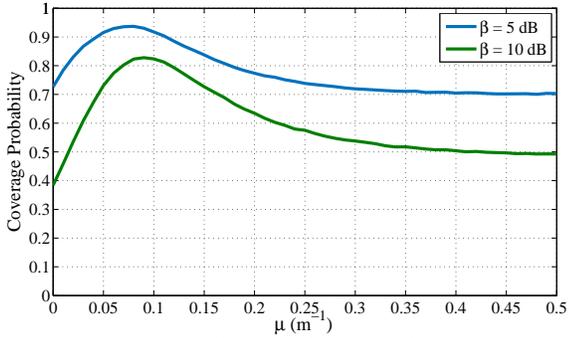}
\vspace{-5pt}
\caption{Coverage probability as a function of blockage exponent $\mu$ with $\delta = \frac{2}{5}$.}
\vspace{-5pt}
\end{figure}
\begin{figure}[t!]
\centering
\center
\vspace{-1ex}
\includegraphics[width =3.4in]{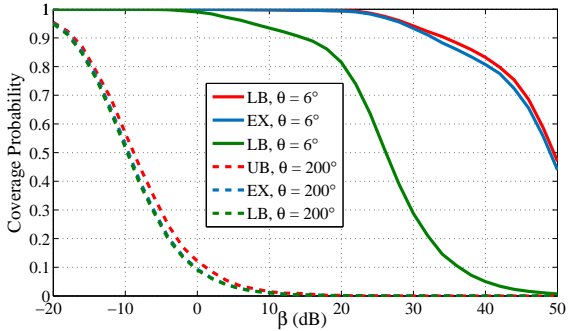}
\vspace{-5pt}
\caption{On the tightness of the coverage probability lower and upper bounds for $\delta = \frac{2}{5}$. EX, LB, and UB denote the exact result, the lower bound, and the upper bound, respectively.}
\vspace{-5pt}
\end{figure}

\begin{figure}[t!]
\centering
\center
\vspace{-1ex}
\includegraphics[width =3.4in]{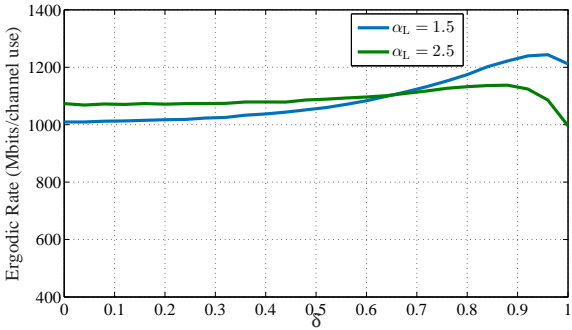}
\vspace{-5pt}
\caption{Spectral efficiency as a function of normalized distance $\delta$.}
\vspace{-7pt}
\end{figure}

\textit{Effect of receiver distance from the center:} The coverage probability as a function of the normalized distance $\delta$ is studied in Fig. 4, considering $\beta = 5$ and $10$ dB. It is observed that, depending on $\beta$, there is an optimal value for the distance of the receiver, about $0.9D$, in terms of the coverage probability. This is due to the fact that the SINR has a tradeoff since the power of both the desired and the interfering LOS and NLOS signals decrease as the distance of the receiver to the center of the disk increases. Also, the transmitters are more likely to be NLOS rather than being LOS.

\textit{Effect of beamwidth:} The coverage probability as a function of the beamwidth $\theta$ is plotted in Fig. 5, considering $\delta = \frac{2}{5}$ and $\beta = 5$ and $10$ dB. As observed, increasing the beamwidth decreases the coverage probability. This is because the main-lobe gain of the antennas in (38) decreases which leads to decreasing the desired power, and also, interfering transmitters are more likely to interfere with the reference receiver with their main antenna beams which leads to increasing the interference power. 

\textit{Effect of blockage exponent:} In Fig. 6, the coverage probability is shown as a function of the blockage exponent $\mu$ for $\delta = \frac{2}{5}$ and $\beta = 5$ and $10$ dB. It is observed, depending on $\beta$, there is an optimal value around 0.075 $\text{m}^{-1}$ for the blockage exponent. That is due to the fact that the SINR has a tradeoff since more transmitters are NLOS as the blockage exponent increases and then the power of both the desired and the interfering signals decrease.

\textit{Tightness of the bounds:} The tightness of the lower and upper bounds on the coverage probability is evaluated in Fig. 7 for $\delta = \frac{2}{5}$ and $\theta = 6^\circ$ and $200^\circ$. As observed, for small $\theta$, i.e., noise-limited networks [30], the upper bound tightly approximate the exact results, while for large $\theta$, i.e., interference-limited networks [30], the lower bound achieves tight results. That is because the upper bound considers minimum interference only from side antenna beams which can be a good approximation when the beamwidth is small. On the other hand, the lower bound considers maximum interference which is the case when transmitters transmit at any direction with their main antenna beams. Also, it is observed that the gap between the lower bound and the exact result for a small $\theta$ is much higher than the the gap between the upper bound and the exact result for a large $\theta$. That is due to the fact that when the transmit beamwidth is small, there is a small probability in alignment of the reference receiver with main antenna beams of transmitters and the main-lobe gain is much higher than the side-lobe gain from (38). On the other hand, according to (38), the difference between the main-lobe gain and the side-lobe gain is small when the beamwidth is large. 

\textit{Ergodic Rate:} The ergodic rate as a function of $\delta$ is shown in Fig. 8 for $\alpha_\text{L} = 1.5$ and $2.5$. As observed, there is an optimal value for the distance of the receiver in terms of the ergodic rate. This is the result of the coverage probability behavior with the distance. Moreover, around $250$ Mbits/channel use and $100$ Mbits/channel use difference in the ergodic rate at the center and the edge of the disk is observed for $\alpha_\text{L} = 1.5$ and $\alpha_\text{L} = 2.5$, respectively, which shows that the location of a receiver plays a key role in its service quality. Also, there is a crossing point, whereby the ergodic rate improves as $\alpha_\text{L}$ increases before reaching a distance for the receiver location. This is because there is a tradeoff since the power of both the desired LOS signal and the interfering LOS signals decrease. 

\section{conclusion}
In this paper, we used stochastic geometry to develop a comprehensive tractable framework for the modeling and analysis of mmwave wireless networks whose nodes are confined in a finite region. We considered a selection strategy to allocate the transmitter with the maximum average received power to a receiver, and accordingly, studied the coverage probability and the ergodic rate over the region. We also proposed upper and lower bounds that are able to tightly approximate the coverage probability at small and large beamwidths, respectively. Our analysis revealed that a higher antenna beamwidth degrades the performance. In addition, according to the setup parameters, there is an optimal blockage exponent and an optimal location for the receiver in terms of the coverage probability and the ergodic rate. 
\appendices
\section{Proof of Theorem 1}
According to (3) and by conditioning on the existence of an LOS or NLOS transmitter inside $\cal A$, the probability that the reference receiver is associated with an LOS transmitter is obtained as 
\begin{align}
{\cal A}_{\text{T},\text{L}}(d) &=\mathbb{P}(n_\text{L} = 0 \hspace{+2pt}\& \hspace{+2pt} n_\text{N} = 0 )\times\nonumber\\&\mathbb{P}\left(\|\mathbf{x}_\text{L}\|^{-\alpha_\text{L}}>\|\mathbf{x}_\text{N}\|^{-\alpha_\text{N}}\mid n_\text{L} = 0 \hspace{+2pt}\& \hspace{+2pt} n_\text{N} = 0\right)\nonumber\hspace{0pt}\\
&+\mathbb{P}(n_\text{L} = 0 \hspace{+2pt}\& \hspace{+2pt} n_\text{N} \geq 1)\times\nonumber\\
&\mathbb{P}\left(\|\mathbf{x}_\text{L}\|^{-\alpha_\text{L}}>\|\mathbf{x}_\text{N}\|^{-\alpha_\text{N}}\mid n_\text{L} = 0 \hspace{+2pt}\& \hspace{+2pt} n_\text{N} \geq 1\right)\nonumber\hspace{0pt}\\&+\mathbb{P}(n_\text{L} \geq 1 \hspace{+2pt}\& \hspace{+2pt} n_\text{N} = 0)\times\nonumber\\
&\mathbb{P}\left(\|\mathbf{x}_\text{L}\|^{-\alpha_\text{L}}>\|\mathbf{x}_\text{N}\|^{-\alpha_\text{N}}\mid n_\text{L} \geq 1 \hspace{+2pt}\& \hspace{+2pt} n_\text{N} = 0\right)\nonumber\hspace{0pt}\\&+\mathbb{P}(n_\text{L} \geq 1 \hspace{+2pt}\& \hspace{+2pt} n_\text{N} \geq 1)\times\nonumber\\&\mathbb{P}\left(\|\mathbf{x}_\text{L}\|^{-\alpha_\text{L}}>\|\mathbf{x}_\text{N}\|^{-\alpha_\text{N}}\mid n_\text{L} \geq 1 \hspace{+2pt}\& \hspace{+2pt} n_\text{N} \geq 1 \right).
\end{align}
Then, due to the following facts
\begin{align}
\left\{\begin{matrix}
&\hspace{-10pt}\mathbb{P}\left(\|\mathbf{x}_\text{L}\|^{-\alpha_\text{L}}>\|\mathbf{x}_\text{N}\|^{-\alpha_\text{N}}\mid n_\text{L} = 0 \hspace{+2pt}\& \hspace{+2pt} n_\text{N} = 0\right) = 0,\\
&\hspace{-10pt}\mathbb{P}\left(\|\mathbf{x}_\text{L}\|^{-\alpha_\text{L}}>\|\mathbf{x}_\text{N}\|^{-\alpha_\text{N}}\mid n_\text{L} = 0 \hspace{+2pt}\& \hspace{+2pt} n_\text{N} \geq 1\right) = 0,\\
&\hspace{-10pt}\mathbb{P}\left(\|\mathbf{x}_\text{L}\|^{-\alpha_\text{L}}>\|\mathbf{x}_\text{N}\|^{-\alpha_\text{N}}\mid n_\text{L} \geq 1 \hspace{+2pt}\& \hspace{+2pt} n_\text{N} = 0\right) = 1,
\end{matrix}\right.
\end{align} 
and 
\begin{align}
\mathbb{P}\left(\|\mathbf{x}_\text{L}\|^{-\alpha_\text{L}}>\|\mathbf{x}_\text{N}\|^{-\alpha_\text{N}}\mid n_\text{L} \geq 1 \hspace{+2pt}\& \hspace{+2pt} n_\text{N} \geq 1 \right)\nonumber\\=\int_{0}^{\infty} \mathbb{P}\left(\|\mathbf{x}_\text{N}\|>r^{\frac{\alpha_\text{L}}{\alpha_\text{N}}}\right) \times f_\text{T,L}^d(r) \mathrm{d}r,
\end{align}
which is obtained by conditioning on the serving distance $r$, and according to (10) and (11) and the facts that $\alpha_\text{N}>\alpha_\text{L}$ and $\mathbb{P}\left(\|\mathbf{x}_\text{N}\|>r^{\frac{\alpha_\text{L}}{\alpha_\text{N}}} \right)$ has crossing points at $(ِD-d)^{\frac{\alpha_\text{N}}{\alpha_\text{L}}}$ and $(D+d)^{\frac{\alpha_\text{N}}{\alpha_\text{L}}}$, we have the following cases to compute (39):

\textbf{Case 1:} If $(ِD-d)^{\frac{\alpha_\text{N}}{\alpha_\text{L}}} > D+d$, we have the order $D-d<D+d< (ِD-d)^{\frac{\alpha_\text{N}}{\alpha_\text{L}}}<(D+d)^{\frac{\alpha_\text{N}}{\alpha_\text{L}}}$, and then by replacing the related values of $\mathbb{P}\left(\|\mathbf{x}_\text{N}\|>r^{\frac{\alpha_\text{L}}{\alpha_\text{N}}}\right)$ and $f_\text{T,L}^d(r)$ for each separate interval, we can write
\begin{align}
{\cal A}_{\text{T},\text{L}}^1(d) = (1-e^{-\lambda_\text{T}{\cal H}_d})e^{-\lambda_\text{T}{\cal G}_d}\times 1+\nonumber\hspace{+90pt}\\
\int_{0}^{D-d} \left(e^{-2 \pi \lambda_\text{T} \int_{0}^{r^{\frac{\alpha_\text{L}}{\alpha_\text{N}}}} x(1-p_\text{L}(x)) \mathrm{d}x}-e^{-\lambda_\text{T}{\cal G}_d}\right)\nonumber\hspace{10pt}\\
\times2\pi \lambda_\text{T} r p_\text{L}(r)e^{-2 \pi \lambda_\text{T} \int_{0}^{r} xp_\text{L}(x) \mathrm{d}x} \mathrm{d}r+\nonumber\hspace{+50pt}\\
\int_{ِD-d}^{D+d}\left(e^{-2 \pi \lambda_\text{T} \int_{0}^{r^{\frac{\alpha_\text{L}}{\alpha_\text{N}}}} x(1-p_\text{L}(x)) \mathrm{d}x}-e^{-\lambda_\text{T}{\cal G}_d}\right) \nonumber\hspace{10pt}\\
\times\lambda_\text{T} \frac{{\partial {\cal H}_d(r) }}{{\partial r}} e^{- \lambda_\text{T} {\cal H}_d(r)}\mathrm{d}r+\nonumber\hspace{+85pt}\\
\int_{D+d}^{(ِD-d)^{\frac{\alpha_\text{N}}{\alpha_\text{L}}}}\left(e^{-2 \pi \lambda_\text{T} \int_{0}^{r^{\frac{\alpha_\text{L}}{\alpha_\text{N}}}} x(1-p_\text{L}(x)) \mathrm{d}x}-e^{-\lambda_\text{T}{\cal G}_d}\right)\times 0 \mathrm{d}r\nonumber\hspace{0pt}\\+\int_{(ِD-d)^{\frac{\alpha_\text{N}}{\alpha_\text{L}}}}^{(D+d)^{\frac{\alpha_\text{N}}{\alpha_\text{L}}}}\left(e^{- \lambda_\text{T} {\cal G}_d\left(r^{\frac{\alpha_\text{L}}{\alpha_\text{N}}}\right)}-e^{-\lambda_\text{T}{\cal G}_d}\right)\times 0\mathrm{d}r\nonumber\\
+\int_{(D+d)^{\frac{\alpha_\text{N}}{\alpha_\text{L}}}}^{\infty}0\times 0\mathrm{d}r.\hspace{110pt}
\end{align}
\textbf{Case 2:} If $(ِD-d)^{\frac{\alpha_\text{N}}{\alpha_\text{L}}} < D+d$, we have the order $D-d< (ِD-d)^{\frac{\alpha_\text{N}}{\alpha_\text{L}}}<ِD+d<(D+d)^{\frac{\alpha_\text{N}}{\alpha_\text{L}}}$, and then by replacing the related values of $\mathbb{P}\left(\|\mathbf{x}_\text{N}\|>r^{\frac{\alpha_\text{L}}{\alpha_\text{N}}}\right)$ and $f_\text{T,L}^d(r)$ for each separate interval, we can write
\begin{align}
{\cal A}_{\text{T},\text{L}}^2(d) &= (1-e^{-\lambda_\text{T}{\cal H}_d})e^{-\lambda_\text{T}{\cal G}_d}\times 1+\nonumber\hspace{+0pt}\\&\int_{0}^{D-d} \left(e^{-2 \pi \lambda_\text{T} \int_{0}^{r^{\frac{\alpha_\text{L}}{\alpha_\text{N}}}} x(1-p_\text{L}(x)) \mathrm{d}x}-e^{-\lambda_\text{T}{\cal G}_d}\right)\nonumber\\&  \times 2\pi \lambda_\text{T} r p_\text{L}(r)e^{-2 \pi \lambda_\text{T} \int_{0}^{r} xp_\text{L}(x) \mathrm{d}x} \mathrm{d}r+\nonumber\hspace{+0pt}\\&\int_{ِD-d}^{(ِD-d)^{\frac{\alpha_\text{N}}{\alpha_\text{L}}}}\left(e^{-2 \pi \lambda_\text{T} \int_{0}^{r^{\frac{\alpha_\text{L}}{\alpha_\text{N}}}} x(1-p_\text{L}(x)) \mathrm{d}x}-e^{-\lambda_\text{T}{\cal G}_d}\right) \nonumber\\
&\times \lambda_\text{T} \frac{{\partial {\cal H}_d(r) }}{{\partial r}} e^{- \lambda_\text{T} {\cal H}_d(r)}\mathrm{d}r+\nonumber\\
&\int_{(ِD-d)^{\frac{\alpha_\text{N}}{\alpha_\text{L}}}}^{D+d} \left(e^{- \lambda_\text{T} {\cal G}_d\left(r^{\frac{\alpha_\text{L}}{\alpha_\text{N}}}\right)}-e^{-\lambda_\text{T}{\cal G}_d}\right) \nonumber\\ &\times\lambda_\text{T} \frac{{\partial {\cal H}_d(r) }}{{\partial r}} e^{- \lambda_\text{T} {\cal H}_d(r)}\mathrm{d}r+ \int_{(D+d)^{\frac{\alpha_\text{N}}{\alpha_\text{L}}}}^{\infty}0\times 0 \mathrm{d}r \hspace{+0pt}+\nonumber\hspace{+0pt}\\&\int_{D+d}^{(D+d)^{\frac{\alpha_\text{N}}{\alpha_\text{L}}}}\left(e^{- \lambda_\text{T} {\cal G}_d\left(r^{\frac{\alpha_\text{L}}{\alpha_\text{N}}}\right)}-e^{-\lambda_\text{T}{\cal G}_d}\right)\times 0 \mathrm{d}r.
\end{align}
With some simplifications, (42) and (43) lead to the final results.

\section{Proof of Theorem 3}
The distribution of the serving distance conditioned on the fact that an LOS transmitter is associated to the reference receiver can be obtained as
\begin{eqnarray}
\mathbb{P}\left(\|\mathbf{x}_\text{L}\|>r\mid \|\mathbf{x}_\text{L}\|^{-\alpha_\text{L}}>\|\mathbf{x}_\text{N}\|^{-\alpha_\text{N}}\right) =\nonumber\hspace{30pt}\\ \frac{\mathbb{P}\left( \|\mathbf{x}_\text{L}\|>r , \|\mathbf{x}_\text{L}\|^{-\alpha_\text{L}}>\|\mathbf{x}_\text{N}\|^{-\alpha_\text{N}}\right)}{\mathbb{P}\left(\|\mathbf{x}_\text{L}\|^{-\alpha_\text{L}}>\|\mathbf{x}_\text{N}\|^{-\alpha_\text{N}}\right) },
\end{eqnarray}
where $\mathbb{P}\left(\|\mathbf{x}_\text{L}\|^{-\alpha_\text{L}}>\|\mathbf{x}_\text{N}\|^{-\alpha_\text{N}}\right)$ is the association probability and
\begin{align}
&\mathbb{P}\left( \|\mathbf{x}_\text{L}\|>r , \|\mathbf{x}_\text{L}\|^{-\alpha_\text{L}}>\|\mathbf{x}_\text{N}\|^{-\alpha_\text{N}}\right) = \mathbb{P}(n_\text{L} = 0 \hspace{+2pt}\& \hspace{+2pt} n_\text{N} = 0 )\times\nonumber\\&\mathbb{P}\left(\|\mathbf{x}_\text{L}\|>r , \|\mathbf{x}_\text{L}\|^{-\alpha_\text{L}}>\|\mathbf{x}_\text{N}\|^{-\alpha_\text{N}}\mid n_\text{L} = 0 \hspace{+2pt}\& \hspace{+2pt} n_\text{N} = 0\right)\nonumber\hspace{0pt}\\&+\mathbb{P}(n_\text{L} = 0 \hspace{+2pt}\& \hspace{+2pt} n_\text{N} \geq 1)\times\nonumber\\
&\mathbb{P}\left(\|\mathbf{x}_\text{L}\|>r , \|\mathbf{x}_\text{L}\|^{-\alpha_\text{L}}>\|\mathbf{x}_\text{N}\|^{-\alpha_\text{N}}\mid n_\text{L} = 0 \hspace{+2pt}\& \hspace{+2pt} n_\text{N} \geq 1\right)\nonumber\hspace{0pt}\\&+\mathbb{P}(n_\text{L} \geq 1 \hspace{+2pt}\& \hspace{+2pt} n_\text{N} = 0)\times\nonumber\\
&\mathbb{P}\left(\|\mathbf{x}_\text{L}\|>r , \|\mathbf{x}_\text{L}\|^{-\alpha_\text{L}}>\|\mathbf{x}_\text{N}\|^{-\alpha_\text{N}}\mid n_\text{L} \geq 1 \hspace{+2pt}\& \hspace{+2pt} n_\text{N} = 0\right)\nonumber\hspace{0pt}\\
&+\mathbb{P}(n_\text{L} \geq 1 \hspace{+2pt}\& \hspace{+2pt} n_\text{N} \geq 1)\times\nonumber\\&\mathbb{P}\left(\|\mathbf{x}_\text{L}\|>r , \|\mathbf{x}_\text{L}\|^{-\alpha_\text{L}}>\|\mathbf{x}_\text{N}\|^{-\alpha_\text{N}}\mid n_\text{L} \geq 1 \hspace{+2pt}\& \hspace{+2pt} n_\text{N} \geq 1 \right).
\end{align}
Then, due to the following facts
\begin{align}
\hspace{-5pt}\left\{\begin{matrix}
\mathbb{P}\left(\|\mathbf{x}_\text{L}\|>r , \|\mathbf{x}_\text{L}\|^{-\alpha_\text{L}}>\|\mathbf{x}_\text{N}\|^{-\alpha_\text{N}}\mid n_\text{L} = 0 \hspace{+2pt}\& \hspace{+2pt} n_\text{N} = 0\right) = 0,\hspace{+0pt}\\
\mathbb{P}\left(\|\mathbf{x}_\text{L}\|>r , \|\mathbf{x}_\text{L}\|^{-\alpha_\text{L}}>\|\mathbf{x}_\text{N}\|^{-\alpha_\text{N}}\mid n_\text{L} = 0 \hspace{+2pt}\& \hspace{+2pt} n_\text{N} \geq 1\right) = 0,\hspace{+0pt}\\
\mathbb{P}\left(\|\mathbf{x}_\text{L}\|>r , \|\mathbf{x}_\text{L}\|^{-\alpha_\text{L}}>\|\mathbf{x}_\text{N}\|^{-\alpha_\text{N}}\mid n_\text{L} \geq 1 \hspace{+2pt}\& \hspace{+2pt} n_\text{N} = 0\right) \hspace{22pt}\\= \mathbb{P}(\|\mathbf{x}_\text{L}\|>r),\hspace{+76pt}\\
\mathbb{P}\left(\|\mathbf{x}_\text{L}\|>r , \|\mathbf{x}_\text{L}\|^{-\alpha_\text{L}}>\|\mathbf{x}_\text{N}\|^{-\alpha_\text{N}}\mid n_\text{L} \geq 1 \hspace{+2pt}\& \hspace{+2pt} n_\text{N} \geq 1 \right)\hspace{22pt}\\= \int_{r}^{\infty} \mathbb{P}\left(\|\mathbf{x}_\text{N}\|>x^{\frac{\alpha_\text{L}}{\alpha_\text{N}}}\right) f_\text{T,L}^d(x) \mathrm{d}x,\hspace{0pt}
\end{matrix}\right.
\end{align} 
and using Theorem 1, we have the following cases to compute (44):

\textbf{Case 1:} If $(ِD-d)^{\frac{\alpha_\text{N}}{\alpha_\text{L}}} > D+d$, then $\mathbb{P}\left(\|\mathbf{x}_\text{L}\|^{-\alpha_\text{L}}>\|\mathbf{x}_\text{N}\|^{-\alpha_\text{N}}\right) = {\cal A}_{\text{T},\text{L}}^1(d)$, and
\begin{align}
&\mathbb{P}\left( \|\mathbf{x}_\text{L}\|>r , \|\mathbf{x}_\text{L}\|^{-\alpha_\text{L}}>\|\mathbf{x}_\text{N}\|^{-\alpha_\text{N}}\right) =\nonumber\hspace{0pt}\\&\int_{r}^{D-d} 2\pi \lambda_\text{T} y p_\text{L}(y)e^{{-2 \pi \lambda_\text{T} \left(\int_{0}^{y^{\frac{\alpha_\text{L}}{\alpha_\text{N}}}} x(1-p_\text{L}(x)) \mathrm{d}x+ \int_{0}^{y} xp_\text{L}(x) \mathrm{d}x\right)}} \mathrm{d}y\nonumber\\&+\int_{ِD-d}^{D+d}\lambda_\text{T} \frac{{\partial {\cal H}_d(y) }}{{\partial y}}e^{{-2 \pi \lambda_\text{T} \left(\int_{0}^{y^{\frac{\alpha_\text{L}}{\alpha_\text{N}}}} x(1-p_\text{L}(x)) \mathrm{d}x+\frac{1}{2\pi} {\cal H}_d(y)\right)}}\mathrm{d}y,\hspace{0pt}
\end{align}
if $0<r<D-d$, and
\begin{align}
&\mathbb{P}\left( \|\mathbf{x}_\text{L}\|>r , \|\mathbf{x}_\text{L}\|^{-\alpha_\text{L}}>\|\mathbf{x}_\text{N}\|^{-\alpha_\text{N}}\right) =
\nonumber\hspace{0pt}\\&\int_{r}^{D+d}\lambda_\text{T} \frac{{\partial {\cal H}_d(y) }}{{\partial y}}e^{{-2 \pi \lambda_\text{T} \left(\int_{0}^{y^{\frac{\alpha_\text{L}}{\alpha_\text{N}}}} x(1-p_\text{L}(x)) \mathrm{d}x+\frac{1}{2\pi} {\cal H}_d(y)\right)}}\mathrm{d}y,
\end{align}
if $D-d<r<D+d$, and
\begin{eqnarray}
\mathbb{P}\left( \|\mathbf{x}_\text{L}\|>r , \|\mathbf{x}_\text{L}\|^{-\alpha_\text{L}}>\|\mathbf{x}_\text{N}\|^{-\alpha_\text{N}}\right) =0,
\end{eqnarray}
if $r>D+d$.

\textbf{Case 2:} If $(ِD-d)^{\frac{\alpha_\text{N}}{\alpha_\text{L}}} < D+d$, then $\mathbb{P}\left(\|\mathbf{x}_\text{L}\|^{-\alpha_\text{L}}>\|\mathbf{x}_\text{N}\|^{-\alpha_\text{N}}\right) = {\cal A}_{\text{T},\text{L}}^2(d)$, and
\begin{align}
&\mathbb{P}\left( \|\mathbf{x}_\text{L}\|>r , \|\mathbf{x}_\text{L}\|^{-\alpha_\text{L}}>\|\mathbf{x}_\text{N}\|^{-\alpha_\text{N}}\right) =\nonumber\hspace{0pt}\\
&\int_{r}^{D-d} 2\pi \lambda_\text{T} y p_\text{L}(y)e^{{-2 \pi \lambda_\text{T} \left(\int_{0}^{y^{\frac{\alpha_\text{L}}{\alpha_\text{N}}}} x(1-p_\text{L}(x)) \mathrm{d}x+ \int_{0}^{y} xp_\text{L}(x) \mathrm{d}x\right)}} \mathrm{d}y\nonumber\hspace{0pt}\\&\hspace{-5pt}+\int_{ِD-d}^{(ِD-d)^{\frac{\alpha_\text{N}}{\alpha_\text{L}}}}\lambda_\text{T} \frac{{\partial {\cal H}_d(y) }}{{\partial y}}e^{{-2 \pi \lambda_\text{T} \left(\int_{0}^{y^{\frac{\alpha_\text{L}}{\alpha_\text{N}}}} x(1-p_\text{L}(x)) \mathrm{d}x+\frac{1}{2\pi} {\cal H}_d(y)\right)}}\mathrm{d}y\hspace{+0pt}\nonumber\\&+\int_{(ِD-d)^{\frac{\alpha_\text{N}}{\alpha_\text{L}}}}^{D+d} \lambda_\text{T} \frac{{\partial {\cal H}_d(y) }}{{\partial y}} e^{{- \lambda_\text{T} \left({\cal G}_d\left(y^{\frac{\alpha_\text{L}}{\alpha_\text{N}}}\right) + {\cal H}_d(y)\right)}}\mathrm{d}y,\hspace{0pt}
\end{align}
if $0<r<D-d$, and
\begin{align}
&\mathbb{P}\left( \|\mathbf{x}_\text{L}\|>r , \|\mathbf{x}_\text{L}\|^{-\alpha_\text{L}}>\|\mathbf{x}_\text{N}\|^{-\alpha_\text{N}}\right)= \nonumber\hspace{0pt}\\&\int_{r}^{(ِD-d)^{\frac{\alpha_\text{N}}{\alpha_\text{L}}}}\lambda_\text{T} \frac{{\partial {\cal H}_d(y) }}{{\partial y}}e^{{-2 \pi \lambda_\text{T} \left(\int_{0}^{y^{\frac{\alpha_\text{L}}{\alpha_\text{N}}}} x(1-p_\text{L}(x)) \mathrm{d}x+\frac{1}{2\pi} {\cal H}_d(y)\right)}}\mathrm{d}y\hspace{+0pt}\nonumber\\&+\int_{(ِD-d)^{\frac{\alpha_\text{N}}{\alpha_\text{L}}}}^{D+d} \lambda_\text{T} \frac{{\partial {\cal H}_d(y) }}{{\partial y}} e^{{- \lambda_\text{T} \left({\cal G}_d\left(y^{\frac{\alpha_\text{L}}{\alpha_\text{N}}}\right) + {\cal H}_d(y)\right)}}\mathrm{d}y,\hspace{0pt}
\end{align}
if $D-d<r<(ِD-d)^{\frac{\alpha_\text{N}}{\alpha_\text{L}}}$, and
\begin{align}
&\mathbb{P}\left( \|\mathbf{x}_\text{L}\|>r , \|\mathbf{x}_\text{L}\|^{-\alpha_\text{L}}>\|\mathbf{x}_\text{N}\|^{-\alpha_\text{N}}\right) =\nonumber\\&\int_{r}^{D+d} \lambda_\text{T} \frac{{\partial {\cal H}_d(y) }}{{\partial y}}e^{{- \lambda_\text{T} \left({\cal G}_d\left(y^{\frac{\alpha_\text{L}}{\alpha_\text{N}}}\right) + {\cal H}_d(y)\right)}}\mathrm{d}y,
\end{align}
if $(ِD-d)^{\frac{\alpha_\text{N}}{\alpha_\text{L}}}<r<D+d$, and
\begin{eqnarray}
\mathbb{P}\left( \|\mathbf{x}_\text{L}\|>r , \|\mathbf{x}_\text{L}\|^{-\alpha_\text{L}}>\|\mathbf{x}_\text{N}\|^{-\alpha_\text{N}}\right) =0,
\end{eqnarray}
if $r>D+d$.
Therefore, the PDF is obtained by taking derivation from the CDF, which is equal to $1-\mathbb{P}(\|\mathbf{x}_\text{L}\|>r\mid \|\mathbf{x}_\text{L}\|^{-\alpha_\text{L}}>\|\mathbf{x}_\text{N}\|^{-\alpha_\text{N}})$.

\section{Characterization of $G_\mathbf{y}$}
We can consider the following cases for the directions of the reference receiver and the transmitter at $\mathbf{y}$ and accordingly find the distribution of $G_\mathbf{y}$.

\textbf{Case 1:} If $0<R_\text{T}(\mathbf{y})<D-\hat d (\mathbf{y}) $ and $0<r<D-d$, then due to the rotation invariancy of the PPP and the fact that $\mathbf{b}(\mathbf{y},R_\text{T}(\mathbf{y}))$ and $\mathbf{b}(\mathbf{o},r)$ are completely inside $\cal A$, the main antenna beams of both the reception at the origin and transmission at $\mathbf{y}$ can have directions with uniform distribution over $2\pi$. Thus, $G_\mathbf{y}$ takes
\begin{eqnarray}
 \left\{\begin{matrix}
a_1 \hspace{+2pt} \text{with prob.}\hspace{+2pt} b_1(\mathbf{y},R_\text{T}(\mathbf{y}),r) = \frac{\theta_\text{T} }{{2\pi }}\frac{\theta_\text{R} }{{2\pi }},\hspace{60pt}\\ 
a_2  \hspace{+2pt} \text{with prob.}\hspace{+2pt} b_2(\mathbf{y},R_\text{T}(\mathbf{y}),r) = \frac{\theta_\text{T} }{{2\pi }}\left(1-\frac{\theta_\text{R} }{{2\pi }}\right),\hspace{30pt}\\
a_3  \hspace{+2pt} \text{with prob.}\hspace{+2pt} b_3(\mathbf{y},R_\text{T}(\mathbf{y}),r) = \left(1-\frac{\theta_\text{T} }{{2\pi}}\right)\frac{\theta_\text{R}}{{2\pi }},\hspace{31pt}\\ 
a_4  \hspace{+2pt} \text{with prob.}\hspace{+2pt} b_4(\mathbf{y},R_\text{T}(\mathbf{y}),r) = \left(1-\frac{\theta_\text{T} }{{2\pi }}\right)\left(1-\frac{\theta_\text{R}}{{2\pi }}\right).\hspace{3pt}\\
 \end{matrix}\right.
\end{eqnarray}

\textbf{Case 2:} If $D-\hat d (\mathbf{y}) <R_\text{T}(\mathbf{y})<D+\hat d (\mathbf{y})$ and $0<r<D-d$, then the main antenna beam of the reception at the origin can have a direction with uniform distribution over $2\pi$. However, since receivers are outside the disk $\mathbf{b}(\mathbf{y},R_\text{T}(\mathbf{y}))$ which intersects with $\cal A$ at angle $\phi_\text{T} (\mathbf{y},R_\text{T}(\mathbf{y}))= {\cos ^{ - 1}}\left( {\frac{{{R_\text{T}(\mathbf{y})^2} + {{\hat d (\mathbf{y}})^2} - {D^2}}}{{2{\hat d (\mathbf{y}})R_\text{T}(\mathbf{y})}}} \right)$ entangled between the line crossing $\mathbf{y}$ and one of the intersection points and the line crossing $\mathbf{y}$ and $\mathbf{x}_\mathbf{o}$, the main antenna beam of the transmission can have a direction with uniform distribution over $2\phi_\text{T}(\mathbf{y},R_\text{T}(\mathbf{y}))$, with an angle between $-\phi_\text{T}(\mathbf{y},R_\text{T}(\mathbf{y}))$ and $\phi_\text{T}(\mathbf{y},R_\text{T}(\mathbf{y}))$. On the other hand, the receiver at the origin is included in the main antenna beam when the main beam has a direction with an angle between $\hat \phi_\text{T} (\mathbf{y})-\frac{\theta_\text{T}}{2}$ and $\hat \phi_\text{T} (\mathbf{y}) +\frac{\theta_\text{T}}{2}$, where $\hat \phi_\text{T} (\mathbf{y}) = \cos^{-1}\left(\frac{{\hat d} (\mathbf{y})^2+\|\mathbf{y}\|^2-d^2}{2{\hat d} (\mathbf{y})\|\mathbf{y}\|}\right)$ is the angle entangled between the line crossing $\mathbf{y}$ and the origin and the line crossing $\mathbf{y}$ and $\mathbf{x}_\mathbf{o}$. Then, dividing the possible event range of the direction to its total range, the probability of having the origin in the main beam of the transmission at $\mathbf{y}$ is $c(\mathbf{y},R_\text{T}(\mathbf{y}),r) =\max\left\{\frac{\min\left\{\hat \phi_\text{T} (\mathbf{y})+\frac{\theta_\text{T}}{2},\phi_\text{T} (\mathbf{y},R_\text{T}(\mathbf{y}))\right\}-\max\left\{\hat \phi_\text{T} (\mathbf{y})-\frac{\theta_\text{T}}{2},-\phi_\text{T} (\mathbf{y},R_\text{T}(\mathbf{y}))\right\}}{2\phi_\text{T} (\mathbf{y},R_\text{T}(\mathbf{y}))},0\right\}$. Please note that when $\min\left\{\hat \phi_\text{T} (\mathbf{y})+\frac{\theta_\text{T}}{2},\phi_\text{T} (\mathbf{y},R_\text{T}(\mathbf{y}))\right\}<\max\left\{\hat \phi_\text{T} (\mathbf{y})-\frac{\theta_\text{T}}{2},-\phi_\text{T} (\mathbf{y},R_\text{T}(\mathbf{y}))\right\}$, the origin cannot be in direction of any possible main antenna beam from $\mathbf{y}$. 

Thus, $G_\mathbf{y}$ takes
\begin{align}
 \left\{\begin{matrix}
a_1 \hspace{+2pt} \text{with prob.}\hspace{+2pt} b_1(\mathbf{y},R_\text{T}(\mathbf{y}),r) = c(\mathbf{y},R_\text{T}(\mathbf{y}),r)\frac{\theta_\text{R}}{{2\pi }},\hspace{60pt}\\ 
a_2 \hspace{+2pt} \text{with prob.}\hspace{+2pt} b_2(\mathbf{y},R_\text{T}(\mathbf{y}),r) = c(\mathbf{y},R_\text{T}(\mathbf{y}),r)\left(1-\frac{\theta_\text{R} }{{2\pi }}\right),\hspace{29pt}\\
a_3 \hspace{+2pt} \text{with prob.}\hspace{+2pt} b_3(\mathbf{y},R_\text{T}(\mathbf{y}),r) = \left(1-c(\mathbf{y},R_\text{T}(\mathbf{y}),r)\right)\frac{\theta_\text{R} }{{2\pi }},\hspace{31pt}\\ 
a_4  \hspace{+2pt}\text{with prob.}\hspace{+2pt} b_4(\mathbf{y},R_\text{T}(\mathbf{y}),r) =\hspace{133pt}\\ \left(1-c(\mathbf{y},R_\text{T}(\mathbf{y}),r)\right)\left(1-\frac{\theta_\text{R} }{{2\pi }}\right),\hspace{3pt}\\
 \end{matrix}\right.
\end{align}

\textbf{Case 3:} If $0<R_\text{T}(\mathbf{y})<D-\hat d (\mathbf{y}) $ and $D-d<r<D+d$, then the main antenna beam of the transmission at $\mathbf{y}$ can have a direction with uniform distribution over $2\pi$. However, since transmitters are outside the disk $\mathbf{b}(\mathbf{o},r)$ which intersects with $\cal A$ at angle $\varphi(r)$ entangled between the line crossing the origin and one of the intersection points and the line crossing the origin and $\mathbf{x}_\mathbf{o}$, the main antenna beam of the reception can have a direction with uniform distribution over $2\varphi(r)$, with an angle between $-\varphi(r)$ and $\varphi(r)$. On the other hand, the transmitter at $\mathbf{y}$ is included in the main antenna beam when the main beam has a direction with an angle between $\hat \phi_\text{R} (\mathbf{y})-\frac{\theta_\text{R}}{2}$ and $\hat \phi_\text{R} (\mathbf{y}) +\frac{\theta_\text{R}}{2}$, where $\hat \phi_\text{R} (\mathbf{y}) = \cos^{-1}\left(\frac{d^2+\|\mathbf{y}\|^2-{\hat d} (\mathbf{y})^2}{2d\|\mathbf{y}\|}\right)$ is the angle entangled between the line crossing $\mathbf{y}$ and the origin and the line crossing the origin and $\mathbf{x}_\mathbf{o}$. Then, dividing the possible event range of the direction to its total range, the probability of having $\mathbf{y}$ in the main beam of the reception at the origin is $d(\mathbf{y},r) =\max\left\{\frac{\min\left\{\hat \phi_\text{R} (\mathbf{y})+\frac{\theta_\text{R}}{2},\varphi (r)\right\}-\max\left\{\hat \phi_\text{R} (\mathbf{y})-\frac{\theta_\text{R}}{2},-\varphi (r)\right\}}{2\varphi (r)},0\right\}$. Please note that when $\min\left\{\hat \phi_\text{R} (\mathbf{y})+\frac{\theta_\text{R}}{2},\varphi (r)\right\}<\max\left\{\hat \phi_\text{R} (\mathbf{y})-\frac{\theta_\text{R}}{2},-\varphi (r)\right\}$, $\mathbf{y}$ cannot be in direction of any possible main antenna beam from the origin. 

Thus, $G_\mathbf{y}$ takes
\begin{align}
 \left\{\begin{matrix}
a_1 \hspace{+2pt} \text{with prob.}\hspace{+2pt} b_1(\mathbf{y},R_\text{T}(\mathbf{y}),r) = \frac{\theta_\text{T} }{{2\pi}}d(\mathbf{y},r),\hspace{60pt}\\ 
a_2  \hspace{+2pt} \text{with prob.}\hspace{+2pt} b_2(\mathbf{y},R_\text{T}(\mathbf{y}),r) = \frac{\theta_\text{T} }{{2\pi}}\left(1-d(\mathbf{y},r)\right),\hspace{31pt}\\
a_3  \hspace{+2pt}\text{with prob.}\hspace{+2pt} b_3(\mathbf{y},R_\text{T}(\mathbf{y}),r) = \left(1-\frac{\theta_\text{T} }{{2\pi }}\right)d(\mathbf{y},r),\hspace{31pt}\\ 
a_4 \hspace{+2pt} \text{with prob.}\hspace{+2pt} b_4(\mathbf{y},R_\text{T}(\mathbf{y}),r) = \left(1-\frac{\theta_\text{T}}{{2\pi }}\right)\left(1-d(\mathbf{y},r)\right),\hspace{4pt}
 \end{matrix}\right.
\end{align}

\textbf{Case 4:} If $D-\hat d (\mathbf{y}) <R_\text{T}(\mathbf{y})<D+\hat d (\mathbf{y}) $ and $D-d<r<D+d$, then the main antenna beams of the transmission and reception can have directions with uniform distribution over $2\phi_\text{T}(\mathbf{y},R_\text{T}(\mathbf{y}))$ and $2\varphi(r)$, respectively. Thus, according to Cases 2 and 3, $G_\mathbf{y}$ takes
\begin{align}
 \left\{\begin{matrix}
a_1 \hspace{2pt}   \text{with prob.}\hspace{+2pt} b_1(\mathbf{y},R_\text{T}(\mathbf{y}),r) =c(\mathbf{y},R_\text{T}(\mathbf{y}),r) d(\mathbf{y},r),\hspace{38pt}\\ 
a_2  \hspace{2pt} \text{with prob.}\hspace{+2pt} b_2(\mathbf{y},R_\text{T}(\mathbf{y}),r) =c(\mathbf{y},R_\text{T}(\mathbf{y}),r)\left(1-d(\mathbf{y},r)\right),\hspace{11pt}\\
a_3   \hspace{2pt} \text{with prob.}\hspace{+2pt} b_3(\mathbf{y},R_\text{T}(\mathbf{y}),r) = \left(1-c(\mathbf{y},R_\text{T}(\mathbf{y}),r)\right)d(\mathbf{y},r),\hspace{12pt}\\ 
a_4  \hspace{2pt} \text{with prob.}\hspace{+2pt} b_4(\mathbf{y},R_\text{T}(\mathbf{y}),r) =\hspace{129pt}\\\left(1-c(\mathbf{y},R_\text{T}(\mathbf{y}),r)\right)\left(1-d(\mathbf{y},r)\right).\hspace{2pt}\\
 \end{matrix}\right.
\end{align}


\begin{thebibliography}{1}
\bibitem{ref1}
J. G. Andrews, S. Buzzi, W. Choi, S. Hanly, A. Lozano, A. C. K. Soong, and J. C. Zhang, "What will 5G be?," \emph{IEEE J. Sel. Areas Commun.}, vol. 32, no. 6, pp. 1065-1082, Jun. 2014.
\bibitem{ref2}
T. S. Rappaport, R. W. Heath, R. C. Daniels, and J. N. Murdock, \emph{Millimeter Wave Wireless Communications}. Pearson Education, 2014.
\bibitem{ref3}
S. Rangan, T. S. Rappaport, and E. Erkip, "Millimeter-wave cellular wireless networks: Potentials and challenges," \emph{Proc. IEEE}, vol. 102, no. 3, pp. 366-385, Mar. 2014.
\bibitem{ref4}
M. Haenggi, J. G. Andrews, F. Baccelli, O. Dousse, and M. Franceschetti, "Stochastic geometry and random graphs for the analysis and design of wireless networks," \emph{IEEE J. Sel. Areas Commun.}, vol. 27, no. 7, pp. 1029-1046, Sep. 2009.
\bibitem{ref5}
J. G. Andrews, F. Baccelli, and R. K. Ganti, "A tractable approach to coverage and rate in cellular networks," \emph{IEEE Trans. Commun.}, vol. 59, no. 11, pp. 3122-3134, Nov. 2011.
\bibitem{ref6}
M. Haenggi, \emph{Stochastic Geometry for Wireless Networks}. Cambridge University Press, 2012.
\bibitem{ref7}
M. Akdeniz, Y. Liu, M. Samimi, S. Sun, S. Rangan, T. Rappaport, and E. Erkip, "Millimeter wave channel modeling and cellular capacity evaluation," \emph{IEEE J. Sel. Areas Commun.}, vol. 32, no. 6, pp. 1164-1179, June 2014.
\bibitem{ref8}
J. G. Andrews, T. Bai, M. N. Kulkarni, A. Alkhateeb, A. K. Gupta, and R. W. Heath, "Modeling and Analyzing Millimeter Wave Cellular Systems," \emph{IEEE Trans. on Commun.}, vol. 65, no. 1, pp. 403-430, Jan. 2017.
\bibitem{ref9}
T. Bai and R. W. Heath, "Coverage and Rate Analysis for Millimeter-Wave Cellular Networks," \emph{IEEE Trans. Wireless Commun.}, vol. 14, no. 2, pp. 1100-1114, Feb. 2015.
\bibitem{ref10}
A. Thornburg, T. Bai, and R. W. Heath, "Performance Analysis of Outdoor mmWave Ad Hoc Networks," \emph{IEEE Trans. Signal Process.}, vol. 64, no. 15, pp. 4065-4079, Aug. 2016.
\bibitem{ref11}
E. Turgut and M. C. Gursoy, "Coverage in Heterogeneous Downlink Millimeter Wave Cellular Networks," \emph{IEEE Trans. Commun.}, vol. 65, no. 10, pp. 4463-4477, May 2017.
\bibitem{ref12}
A. K. Gupta, A. Alkhateeb, J. G. Andrews, and R. W. Heath, "Gains of Restricted Secondary Licensing in Millimeter Wave Cellular Systems," \emph{IEEE J. Sel. Areas Commun.}, vol. 34, no. 11, pp. 2935-2950, Nov. 2016.
\bibitem{ref13}
X. Yu, J. Zhang, M. Haenggi, and K. B. Letaief, "Coverage Analysis for Millimeter Wave Networks: The Impact of Directional Antenna Arrays," \emph{IEEE J. Sel. Areas Commun.}, vol. 35, no. 7, pp. 1498-1512, July 2017.
\bibitem{ref14}
N. Deng and M. Haenggi, "A Fine-Grained Analysis of Millimeter-Wave Device-to-Device Networks", \emph{IEEE Trans. on Commun.}, vol. 65, no. 11, pp. 4940-4954, Nov. 2017.
\bibitem{ref15}
C. Park and T. Rappaport, "Short-range wireless communications for next-generation networks: UWB, 60 GHz millimeter-wave WPAN, and zigbee," \emph{IEEE Wireless Commun.}, vol. 14, pp. 70-78, Aug. 2007.
\bibitem{ref16}
WirelessHD, "WirelessHD specification overview," 2010. [Online]. Available: http://www.wirelesshd.org
\bibitem{ref17}
E. Perahia, C. Cordeiro, M. Park, and L. L. Yang, "IEEE 802.11ad: Defining the next generation multi-Gbps Wi-Fi,"
\emph{Proc. IEEE Consumer Commun. and Networking Conf.}, Las Vegas, USA, pp. 1-5, Jan. 2010.
\bibitem{ref18}
S. M. Azimi-Abarghouyi, B. Makki, M. Haenggi, M. Nasiri-Kenari, and T. Svensson, "Stochastic Geometry Modeling and Analysis of Single- and Multi-Cluster Wireless Networks," \emph{IEEE Trans. Commun.}, to appear, May 2018. [Online]. Available: https://arxiv.org/abs/1712.08784
\bibitem{ref19}
S. M. Azimi-Abarghouyi, B. Makki, M. Haenggi, M. Nasiri-Kenari, and T. Svensson, "Coverage Analysis of Finite Cellular Networks: A Stochastic Geometry Approach," \emph{Iran Workshop on Commun. and Inf. Theory}, Tehran, Iran, April 2018.
\bibitem{ref20}
M. Afshang, and H. S. Dhillon, "Fundamentals of Modeling Finite Wireless Networks using Binomial Point Process", \emph{IEEE Trans. Wireless Commun.}, vol. 16, no. 5, pp. 3355-3370, May 2017. 

\bibitem{ref21}
K. Venugopal, M. C. Valenti, and R. W. Heath, "Interference in finite-sized highly dense millimeter wave networks," \emph{IEEE ITA'2015}, San Diego, USA, 2015, pp. 175-180.
\bibitem{ref22}
K. Venugopal, M. C. Valenti, and R. W. Heath, "Device-to-Device Millimeter Wave Communications: Interference, Coverage, Rate, and Finite Topologies," \emph{IEEE Trans. Wireless Commun.}, vol. 15, no. 9, pp. 6175-6188, Sep. 2016.
\bibitem{ref23}
G. George, K. Venugopal, A. Lozano, and R. W. Heath, "Enclosed mmWave Wearable Networks: Feasibility and Performance," \emph{IEEE Trans. Wireless Commun.}, vol. 16, no. 4, pp. 2300-2313, April 2017.

\bibitem{ref24}
T. Rappaport, G. Maccartney, M. Samimi, and S. Sun, "Wideband millimeter-wave propagation measurements and channel models
for future wireless communication system design," \emph{IEEE Trans. Commun.}, vol. 63, no. 9, pp. 3029-3056, Sept. 2015.
\bibitem{ref25}
H. Alzer, "On some inequalities for the incomplete Gamma function," \emph{Mathematics of Computation}, vol. 66, no. 218, pp. 771-778, 1997.
\bibitem{ref26}
T. D. Novlan, H. S. Dhillon, and J. G. Andrews, "Analytical Modeling of Uplink Cellular Networks," \emph{IEEE Trans. Wireless Commun.}, vol. 12, no. 6, pp. 2669-2679, June 2013.
\bibitem{ref27}
O. Onireti, A. Imran, and M. A. Imran, "Coverage, Capacity and Energy Efficiency Analysis in the Uplink of mmWave Cellular Networks," \emph{IEEE Trans. Veh. Technol.}, vol. 67, no. 5, pp. 3982-3997, May 2018.
\bibitem{ref28}
M. D. Renzo and P. Guan, "Stochastic Geometry Modeling and System-Level Analysis of Uplink Heterogeneous Cellular Networks with Multi-Antenna Base Stations," \emph{IEEE Trans. Commun.}, vol. 64, no. 6, pp. 245-2476, June 2016.
\bibitem{ref29}
H. ElSawy and E. Hossain, "On Stochastic Geometry Modeling of Cellular Uplink Transmission With Truncated Channel Inversion Power Control," \emph{IEEE Trans. Commun.}, vol. 13, no. 8, pp. 4454-4469, Aug. 2014.
\bibitem{ref30}
H. Shokri-Ghadikolaei and C. Fischione, "Millimeter wave ad hoc networks: Noise-limited or interference-limited?," \emph{IEEE
GLOBECOM' 15}, San Diego, USA, Dec. 2015.
\end{thebibliography}
\end{document}